\documentclass[preprintnumbers, prd, twocolumn, showpacs, floatfix, preprintnumbers, 
letterpaper, 
superscriptaddress,nofootinbib]{revtex4-2}
\usepackage{amsmath}
\usepackage{amsfonts}
\usepackage{amssymb}
\usepackage{mathtools}
\usepackage{bm}
\usepackage{xcolor}
\usepackage{hyperref}
\usepackage{physics}
\usepackage[maxfloats=256]{morefloats}
\usepackage{graphicx}

\usepackage{color}
\usepackage{relsize}

\newcommand{\be}{\begin{equation}}
\newcommand{\ee}{\end{equation}}
\newcommand{\bea}{\begin{eqnarray}}
\newcommand{\eea}{\end{eqnarray}}

\begin{document}

\title{ Cosmological Dynamics in Interacting Scalar-Torsion $f(T,\phi)$ Gravity: Investigating Energy and Momentum Couplings }

\author{Carlos Rodriguez-Benites}
\email{cerodriguez@unitru.edu.pe}
\affiliation{Departamento Acad\'emico de F\'{\i}sica, Facultad de Ciencias F\'{\i}sicas y Matem\'aticas, Universidad Nacional de Trujillo, Av. Juan Pablo II s/n, Trujillo, Per\'u}
\affiliation{GRACOCC \& OASIS research groups, Facultad de Ciencias F\'{\i}sicas y Matem\'aticas, Universidad Nacional de Trujillo, Av. Juan Pablo II s/n, Trujillo, Per\'u}

\author{Manuel Gonzalez-Espinoza}
\email{manuel.gonzalez@pucv.cl}
\affiliation{Instituto de F\'{\i}sica, Pontificia Universidad Cat\'olica de 
Valpara\'{\i}so, 
Casilla 4950, Valpara\'{\i}so, Chile}

\author{Giovanni Otalora}
\email{giovanni.otalora@academicos.uta.cl}
\affiliation{Departamento de F\'isica, Facultad de Ciencias, Universidad de Tarapac\'a, Casilla 7-D, Arica, Chile}

\author{Manuel Alva-Morales}
\email{malvam@unitru.edu.pe}
\affiliation{GRACOCC \& OASIS research groups, Facultad de Ciencias F\'{\i}sicas y Matem\'aticas, Universidad Nacional de Trujillo, Av. Juan Pablo II s/n, Trujillo, Per\'u}
\affiliation{Escuela Profesional de F\'{\i}sica, Facultad de Ciencias F\'{\i}sicas y Matem\'aticas, Universidad Nacional de Trujillo, Av. Juan Pablo II s/n, Trujillo, Per\'u}

\date{\today}

\begin{abstract} 
We investigate the cosmological dynamics of a homogeneous scalar field non-minimally coupled to torsion gravity, which also interacts with cold dark matter through energy and momentum transfer. The matter and radiation perfect fluids are modeled using the Sorkin-Schutz formalism. We identify scaling regimes of the field during both the radiation and matter eras. Additionally, we discovered a field-dominated scaling attractor; however, it does not exhibit accelerated expansion, making it unsuitable for describing dark energy. Nevertheless, we find two attractor solutions that do exhibit accelerated expansion: one is a quintessence-like fixed point, and the other is a de Sitter fixed point.

\end{abstract}


\maketitle

\section{Introduction}\label{Introduction}

In 1998, the analysis of Type Ia supernova (SNIa) data revealed that our Universe is expanding at an accelerating rate \cite{Riess:1998cb, Perlmutter:1998np}. Despite this discovery, there is no definitive explanation for this phenomenon. The prevailing interpretation attributes this accelerated expansion to dark energy, which could be a new form of exotic matter or a modification to gravity. Dark energy is believed to constitute 68\% of the Universe's matter-energy density \cite{Aghanim:2018eyx, Ade:2015rim}. While standard cosmology based on Einstein's General Relativity has successfully explained this acceleration through the cosmological constant $\Lambda$, the $\Lambda$CDM model (cosmological constant $\Lambda$ and cold dark matter) faces a severe fine-tuning problem related to its energy scale \cite{Bull:2015stt, Martin:2012bt, Copeland:2006wr, amendola2010dark}. Additionally, recent analyses have identified statistically significant tensions within the $\Lambda$CDM model, such as the $H_0$ discrepancy between the cosmic microwave background (CMB) measurements and direct local distance ladder measurements \cite{Riess:2011yx, Riess:2016jrr, Riess:2018byc, DiValentino:2020zio}, as well as tensions involving the matter energy density $\Omega_{m}$ and the structure growth rate ($f\sigma 8$) \cite{Hildebrandt:2016iqg, Kuijken:2015vca, Conti:2016gav, DiValentino:2018gcu, DiValentino:2020vvd}. {A detailed analysis of the discrepancy between current data and the standard cosmological model is shown in \cite{McGaugh:2023nkc,Battye:2014qga,DES:2021wwk}, showing the tension present in the mass density as well as the Hubble constant.} These tensions suggest the need for investigating new physics beyond the standard cosmological model \cite{Riess:2019cxk, Davari:2019tni, DiValentino:2015bja, Sola:2019jek, Sola:2020lba, Joyce:2014kja, Koyama:2015vza}.

A promising alternative to explain dark energy is through scalar fields, which have been extensively studied in the literature. This includes models such as quintessence \cite{Wetterich:1987fm, Ratra:1987rm, Carroll:1998zi, Tsujikawa:2013fta}, k-essence \cite{Chiba:1999ka, ArmendarizPicon:2000dh, ArmendarizPicon:2000ah}, and tachyon fields \cite{Sen:2002nu, Sen:2002in}, among others \cite{Copeland:2006wr, amendola2010dark}. From the perspective of quantum field theory in curved spacetime, a non-minimal coupling to gravity can naturally arise through quantum corrections \cite{Linde:1982zj} or renormalizability requirements \cite{Freedman:1974gs, Freedman:1974ze, Birrell:1982ix}. For example, the extended quintessence model, which involves a quintessence field coupled to gravity, was first proposed in \cite{Perrotta:1999am} and further explored in \cite{Sahni:1998at, Chiba:1999wt, Bartolo:1999sq, Faraoni:2000wk, Hrycyna:2008gk, Hrycyna:2007gd}. Similarly, non-minimally coupled k-essence and tachyonic fields have been investigated in \cite{Sen:2008bg} and \cite{deSouza:2008nj}, respectively. Galileon models also benefit from a non-minimal coupling to curvature, which helps avoid pathological instabilities or the propagation of additional degrees of freedom \cite{Deffayet:2009wt}. Recent studies have shown that non-minimally coupled scalar field theories can alleviate current observational tensions within the concordance model \cite{Davari:2019tni, DiValentino:2019jae}.

Teleparallel Gravity (TG) offers an equivalent description of gravity through torsion rather than curvature \cite{Einstein, TranslationEinstein, Early-papers1, Early-papers2, Early-papers3, Early-papers4, Early-papers5, Early-papers6, JGPereira2, AndradeGuillenPereira-00, Arcos:2005ec, Pereira:2019woq}. In TG, the tetrad fields replace the metric tensor, and the Weitzenböck connection replaces the Levi-Civita connection \cite{JGPereira2, AndradeGuillenPereira-00, Arcos:2005ec, Pereira:2019woq}. The TG Lagrangian density is proportional to the torsion scalar $T$, which differs from the curvature scalar $R$ by a total derivative term, making the two theories equivalent at the field equations level \cite{Aldrovandi-Pereira-book, Arcos:2005ec}. Extending TG to include a non-minimally coupled scalar-torsion theory leads to models like the one proposed in \cite{Cai:2015emx, Bahamonde:2017ize, Paliathanasis:2022xoq, Paliathanasis:2021nqa, Leon:2023idl, Bamba:2012cp, Bamba:2012vg}, where a scalar field $\phi$ is coupled to the torsion scalar $T$ via a term $\xi \phi^2 T$. This model was first applied to dark energy in \cite{Geng:2011aj, Geng:2011ka} and further extended in \cite{Otalora:2013tba, Otalora:2013dsa}. Unlike the curvature-based theories, these scalar-torsion theories belong to a distinct class of gravitational modifications.

{Recent studies highlight TG’s potential for a renormalizable quantum gravity framework, although divergences arise in one-loop quantum corrections when coupled with matter fields. To address these, modified TG actions have been proposed, with additional free parameters to absorb divergences without destabilizing the theory. This approach entails imposing Lorentz symmetry post-quantization, enabling renormalizability without introducing higher-order derivatives that could generate ghosts \cite{Casadio:2021zai}. Furthermore, extensions incorporating noncommutative geometry through Moyal deformation quantization show promise by introducing a natural cutoff scale for UV divergences, which may lead to a renormalizable TG formulation in extreme regimes, such as near singularities \cite{Matsuyama:2018ywn}.}

{Similarly, scalar-torsion coupling models, such as those including a $\xi \phi^2 T$ term, also support the pursuit of renormalization-compatible modifications in TG. By adding dynamical scalar fields, these models introduce new degrees of freedom that may stabilize high-energy fluctuations \cite{Cai:2015emx, Bahamonde:2017ize}. Alternatively, TG can be applied as an effective field theory limited to scales below the Planck energy, where it remains predictive and finite without full renormalizability constraints \cite{Bajardi:2021tul}. These approaches suggest that TG, particularly through scalar-torsion and noncommutative modifications, may serve as a viable framework for quantization and renormalization of gravity.}

Further generalizations involve introducing terms in the form $F(\phi) G(T)$ in the action, where $G(T)$ is a function of the torsion scalar $T$, or considering a general function $f(T, \phi)$ \cite{Bengochea:2008gz, Linder:2010py, Li:2011wu}. These modifications can also include a non-minimal coupling between the torsion scalar and matter fields \cite{Harko:2014aja, Harko:2014sja, Carloni:2015lsa, Gonzalez-Espinoza:2018gyl}, similar to the curvature-matter coupling in $f(R)$ gravity \cite{Nojiri:2004bi, Allemandi:2005qs, Nojiri:2006ri, Bertolami:2007gv, Harko:2008qz, Harko:2010mv, Bertolami:2009ic, Bertolami:2013kca, Wang:2013fja}. These theories are motivated by counterterms that appear during the quantization of self-interacting scalar fields in curved spacetime \cite{birrell1984quantum}. For example, a generalized scalar-torsion $f(T, \phi)$ gravity theory has been shown to be necessary for explaining primordial fluctuations during slow-roll inflation \cite{Gonzalez-Espinoza:2020azh}. {Evenmore, the stability and bifurcation analyses in $f(R)$ models and $f(T, \phi) $ gravity frameworks have important distinctions due to the nature of the modifications involved. In $ f(R) $ gravity, stability and bifurcations are closely tied to how the function of the Ricci scalar $R $ influences the cosmic expansion phase \cite{Amendola:2006we}. However, in \( f(T, \phi) \) models—where torsion, \( T \), rather than curvature dominates and couples to a scalar field \( \phi \)—the dynamics are driven by an interaction between torsion and scalar fields. This adds complexity to stability conditions since the scalar-torsion coupling introduces additional degrees of freedom, altering fixed points and possible bifurcation structures \cite{Duchaniya:2022fmc}. Moreover, the scalar field in \( f(T, \phi) \) theories can affect the expansion rate in ways not possible within the traditional curvature-based \( f(R) \) frameworks. This often results in new stable configurations or bifurcations that depend on the specific forms of the coupling functions, which do not appear in \( f(R) \) models \cite{Amendola:2006we, Duchaniya:2022fmc, Li:2010cg, Wu:2011xa}. Consequently, by studying \( f(T, \phi) \) models, your work addresses these unique stability aspects, providing a broader perspective on modified gravity theories that consider both torsional and scalar degrees of freedom.} For dark energy at late times, these theories demonstrate new scaling solutions and attractor fixed points with accelerated expansion \cite{Gonzalez-Espinoza:2020jss,Gonzalez-Espinoza:2022hui,Gonzalez-Espinoza:2023whd, Uzan:1999ch, Amendola:1999qq, Copeland:2006wr, amendola2010dark, Kadam:2024vlw, Kadam:2024fgz,Kadam:2023ufk,Duchaniya:2023aeu}.

While these modified gravity theories can account for the observed accelerated expansion at late times and early inflation, they introduce at least one additional degree of freedom. It is crucial to ensure that the evolution of these modes does not result in pathological instabilities such as ghost, Laplacian, or tachyonic instabilities \cite{DeFelice:2016ucp, Heisenberg:2016eld, Kase:2014cwa, DeFelice:2011bh, Sbisa:2014pzo}. At the perturbation level, these additional modes are coupled to those of the matter fields, necessitating a complete stability analysis that includes matter interactions \cite{Gergely:2014rna, Gleyzes:2014qga}. 

Interacting scenarios between dark matter and dark energy appear as candidates to solve or alleviate the cosmological coincidence problem and have been widely studied, mainly through a phenomenological interaction kernel, which has been shown to affect the evolution of the Universe \cite{wang2016dark,wang2024further,rodriguez2024revisiting,rodriguez2020universe,Cid:2020kpp}. The interaction kernel may also provide a momentum exchange between the dark components, and it has been studied in alternative descriptions of the matter sector \cite{amendola2020scaling,jimenez2021velocity,beltran2021probing,jimenez2023smoking,jimenez2020cosmological,asghari2019structure}. In this context, the Sorkin-Schutz action provides a framework to describe the matter sector, allowing for a comprehensive stability analysis in the presence of matter \cite{Schutz:1977df, Brown:1992kc,Gonzalez-Espinoza:2021mwr}. This analysis is essential for determining the viability of the theory before it can be compared with observational data \cite{Heisenberg:2016eld, DeFelice:2016ucp}.

This paper is organized as follows: In section \ref{Intro_TG}, we briefly introduce the basic elements of teleparallel gravity.  In Section \ref{model}, we establish the general action to be studied. We develop the phase space analysis for the FLRW universe, obtaining the critical points and stability conditions. 
In Section \ref{num_a}, we numerically integrate the full cosmological equations using the dynamical analysis approach, corroborating the analytical results obtained in the previous sections. Finally, in Section \ref{Remarks}, we summarize the results obtained.


\section{An Overview of Teleparallel Gravity}\label{Intro_TG}

The Teleparallel Equivalent of General Relativity, also called Teleparallel Gravity (TG), is a gauge theory for the translation group \cite{Early-papers5,Early-papers6,Aldrovandi-Pereira-book,Pereira:2019woq,JGPereira2,Arcos:2005ec}, in which the
dynamical variable is the tetrad field that satisfies
\be
g_{\mu \nu}=\eta_{A B} \ e^{A}_{~\mu} e^{B}_{~\nu}, 
\ee where $g_{\mu \nu}$  is the spacetime metric, and $\eta _{AB}^{}=\text{diag}\,(-1,1,1,1)$ is the Minkowski tangent space metric.  

The Lorentz (or spin) connection of TG is defined as
\be
\omega^{A}_{~B \mu}=\Lambda^{A}_{~D}(x) \partial_{\mu}{\Lambda_{B}^{~D}(x)},
\label{spin_TG}
\ee where $\Lambda^{A}_{~D}(x)$ are the components of a local (point-dependent) Lorentz transformation. For this connection one has a vanishing curvature tensor
\be
R^{A}_{~B \mu\nu}=\partial_{\mu} \omega^{A}_{~B\nu}-\partial_{\nu}{\omega^{A}_{~B \mu}}+\omega^{A}_{~C \mu} \omega^{C}_{~B \nu}-\omega^{A}_{~C \nu} \omega^{C}_{~B \mu}=0, 
\ee while in the presence of gravity this provides us with a non-vanishing torsion tensor
\be
T^{A}_{~~\mu \nu}=\partial_{\mu}e^{A}_{~\nu} -\partial_{\nu}e^{A}_{~\mu}+\omega^{A}_{~B\mu}\,e^{B}_{~\nu}
 -\omega^{A}_{~B\nu}\,e^{B}_{~\mu}.
\ee 
Due to these properties, this connection is called a purely inertial connection or simply flat connection.

Additionally, one can construct a spacetime-indexed linear connection, the so called Weitzenb\"{o}ck connection, in the form
\be
\Gamma^{\rho}_{~~\nu \mu}=e_{A}^{~\rho}\partial_{\mu}e^{A}_{~\nu}+e_{A}^{~\rho}\omega^{A}_{~B \mu} e^{B}_{~\nu}.
\ee 

By introducing the contortion tensor
\begin{equation}  \label{Contortion}
 K^{\rho}_{~~\nu\mu}= \frac{1}{2}\left(T^{~\rho}_{\nu~\mu}
 +T^{~\rho}_{\mu~\nu}-T^{\rho}_{~~\nu\mu}\right),
\end{equation} where $T^{\rho}_{~~\mu \nu}=e_{A}^{~\rho} T^{A}_{~~\mu \nu}$ is the purely spacetime form of the torsion tensor, that the 
Weitzenb\"{o}ck connection satisfies the general relation
\be
\Gamma^{\rho}_{~~\nu \mu}=\bar{\Gamma}^{\rho}_{~~\nu \mu}+K^{\rho}_{~~\nu \mu},
\label{RelGamma}
\ee
where $\bar{\Gamma}^{\rho}_{~~\nu \mu}$ is the known Levi-Civita connection of GR,  and such that 
\be
T^{\rho}_{~~\mu \nu}=\Gamma^{\rho}_{~~\nu \mu}-\Gamma^{\rho}_{~~\mu \nu}.
\ee

Given its foundations as a gauge theory, the action of TG is constructed using quadratic terms in the torsion tensor
\cite{Aldrovandi-Pereira-book}
\be
S=-\frac{1}{2 \kappa^2} \int{d^{4}x e ~T},
\ee where $\kappa^2=8\pi G$, $e=\det{(e^{A}_{~\mu})}=\sqrt{-g}$, and $T$ is the torsion scalar such that
\be
T= S_{\rho}^{~~\mu\nu}\,T^{\rho}_{~~\mu\nu},
\label{ScalarT}
 \ee
with
\begin{equation} \label{Superpotential}
 S_{\rho}^{~~\mu\nu}=\frac{1}{2}\left(K^{\mu\nu}_{~~~\rho}+\delta^{\mu}_{~\rho} \,T^{\theta\nu}_{~~~\theta}-\delta^{\nu}_{~\rho}\,T^{\theta\mu}_{~~~\theta}\right)\,,
\end{equation} the super-potential tensor.  

Putting Eq. \eqref{RelGamma} into \eqref{ScalarT}, one can show that 
\be
T=-R+2 e^{-1} \partial_{\mu}(e T^{\nu \mu}_{~~~\nu}), 
\label{Equiv} 
\ee where $R$ is the curvature scalar of GR. Since $T$ and $R$ differ by a total derivative term, the two theories, TG and GR, are equivalent at the level of field equations. 

Alternatively, modified gravity models can be developed from either curvature-based or torsion-based theories, which may lead to non-equivalent results. In the context of modified teleparallel gravity, various studies have examined dark energy and inflation driven by non-minimally coupled scalar fields \cite{Geng:2011aj,Otalora:2013tba,Otalora:2013dsa,Otalora:2014aoa,Skugoreva:2014ena}.Additionally, models incorporating non-linear torsion terms, like $f(T)$ gravity \cite{Bengochea:2008gz,Linder:2010py}, have been explored \cite{Li:2011wu,Gonzalez-Espinoza:2018gyl}. These torsion-based theories, distinct from curvature-based ones, have led to extensive research in early and late-time cosmology \cite{Cai:2015emx}.

\
\section{Sorkin-Shutz Approach to Matter Fluids}
A general matter fluid composed of barotropic perfect fluids in the 
Sorkin-Schutz formalism, is described by the action \cite{Schutz:1977df, Brown:1992kc, Amendola:2020ldb}
\be
S_{M}=-\sum_{I}\int{d^{4}x\left[e \rho_{I}(n_{I})+J^{\nu}_{I}\partial_{\nu}\ell_{I}\right]},
\label{SS_action}
\ee where $I$ denotes the different species, such as non-relativistic matter ($I=m$, including cold dark matter and baryons), and radiation ($I=r$). Here, $\rho_{I}$ is the energy density, which depends on the number density $n_{I}$, $\ell_{I}$ is a scalar field, and $J^{\nu}_{I}$ is a vector density of weight one. 

The number density $n_{I}$ is defined as
\be
n_{I}=\frac{\sqrt{-J^{\alpha}_{I}J^{\beta}_{I} g_{\alpha \beta}}}{e},
\ee allowing the four-velocity to be expressed as
\be
u^{\alpha}_{I}=\frac{J^{\alpha}_{I}}{n_{I} e},
\label{u_Vel1}
\ee which satisfies the orthogonality relation $g_{\mu \nu} u^{\mu}_{I}u^{\nu}_{I}=-1$ \footnote{It is important to note that, for a homogeneous and isotropic background, the 4-velocity \( u_\mu \) is the same for the matter and radiation eras, $\frac{J^\mu_m}{n_m \sqrt{-g}} = \frac{J^\mu_r}{n_r \sqrt{-g}}$; however, this equivalence holds at the background level and does not necessarily extend to the perturbation level.}. 

The variation of the matter action \eqref{SS_action} with respect to $J^{\alpha}_{I}$ leads us to
\be
u_{I \alpha}=\frac{1}{\rho_{I,n_{I}}}\partial_{\alpha}{\ell_{I}}.
\label{u_Vel2}
\ee where $u_{I \alpha}\equiv g_{\alpha \beta} u^{\beta}_{I}$, and $\rho_{I,n_{I}}\equiv \partial{\rho_{I}}/\partial{n_{I}}$. 

On the other hand, the total matter energy-momentum tensor is given by
\be
T_{\mu}^{~\nu}\equiv e^{A}_{~\mu}\left[\frac{1}{e}\frac{\delta S_{M}}{\delta e^{A}_{~\nu}}\right]=\sum_{I}{T^{(I)\nu}_{\mu}},
\ee
where
\be
T^{(I)\nu}_{\mu}= n_{I}\rho_{I,n_{I}} u_{I \mu}u_{I}^{\nu}+\left(n_{I}\rho_{I,n_{I}}-\rho_{I}\right)\delta^{\nu}_{\mu},
\ee 
which corresponds to the usual energy-momentum tensor of a perfect fluid. Here, the pressure is defined as
\be
p_{I}\equiv n_{I} \rho_{I,n_{I}}-\rho_{I}.
\label{pressure}
\ee
As expected, for $\rho_{I} \propto n_{I}^{1+w_{I}}$, this yields the equation of state $p_{I}=w_{I} \rho_{I}$. 

Varying the matter action with respect to $\ell_{I}$ gives us 
\be
\partial_{\alpha}{J_{I}^{\alpha}}=0.
\label{Conser_Part}
\ee This last equation represents the conservation of particle number for each species in the fluid. By using Eqs. \eqref{u_Vel1} and \eqref{pressure}, Eq. \eqref{Conser_Part} can be written as
\be
u_{I}^{\mu}\partial_{\mu}{\rho_{I}}+\left(\rho_{I}+P_{I}\right)\nabla_{\mu}{u_{I}^{\mu}}=0,
\label{Continuity_Eq}
\ee where $\nabla_{\mu}$ is the covariant derivative operator of the Levi-Civita connection. This is the continuity equation for the energy-momentum tensor of each component. 

Thus, from Eq. \eqref{Continuity_Eq}, one obtains that the perfect-fluid energy-momentum tensor satisfies the conservation law
\be
u_{I\nu}\nabla^{\mu}{T^{(I)\nu}_{\mu}}=-\left[u_{I}^{\mu}\partial_{\mu}{\rho_{I}}+\left(\rho_{I}+P_{I}\right)\nabla_{\mu}{u_{I}^{\mu}}\right]=0,
\ee being equivalent to Eq. \eqref{Conser_Part}. In the case of an isotropic and homogeneous cosmological background, this equation becomes, $u_{\nu}\nabla^{\mu}{T^{(I)\nu}_{\mu}}=0$, where $u_{\nu}$ is the four-velocity, which is the same for all components. For more details, the reader is referred to \cite{amendola2020scaling}.

\section{Interacting scalar-torsion $f(T,\phi)$ gravity}\label{model}

The relevant action is constructed from the scalar-torsion $f(T,\phi)$ gravity action \cite{Hohmann:2018rwf,Gonzalez-Espinoza:2020azh,Gonzalez-Espinoza:2020jss,Gonzalez-Espinoza:2021mwr} by incorporating interactions that involve energy and momentum transfer between the scalar field and cold dark matter \cite{Amendola:2020ldb}. This is achieved using the Sorkin-Schutz formalism \cite{Schutz:1977df, Brown:1992kc}, as follows:
{\small
\bea
 S&=&\int d^{4}x\,e\,\left[f(T,\phi)-f_{1}(\phi,X,Z)\rho_{m}+f_{2}(\phi,X,Z)\right]-\nonumber\\ 
 && \sum_{I=m,r}\int{d^{4}x\left[e \rho_{I}(n_I)+J_I^{\nu}\partial_{\nu}\ell_j\right]},
\label{action_Scalar_Torsion}
\eea} 
where $f(T,\phi)$ is an arbitrary function of the torsion scalar $T$, and the scalar field $\phi$, with $X=-\nabla^\mu\phi\nabla_\mu\phi/2$ as its kinetic term. 
The variable $Z=u^\mu_{m}\nabla_\mu\phi$ is a scalar combination involving the field derivative coupling with the fluid four velocity of dark matter $u^\mu_{m}$ . 

The interaction between the scalar field and cold dark matter fluid is introduced through the functions $f_1$ and $f_2$, both of which depend on $\phi$, $X$, and $Z$. The function $f_1$ provides the energy transfer, while $f_2$ accounts for both momentum exchange and the scalar potential \cite{amendola2020scaling}. 



\subsection{Cosmological dynamics}\label{cosmo_dyna}

In this section, we explore the cosmological dynamics of this model by introducing the cosmological background and useful cosmological parameters. To analyze cosmology within this interacting model, we define the cosmological background by assuming a diagonal tetrad field:
\be
\label{veirbFRW}
e^A_{~\mu}={\rm
diag}(1,a,a,a),
\ee
which corresponds to the Friedmann-Lemaître-Robertson-Walker
(FLRW) spacetime metric with flat spacelike sections
\begin{equation}
ds^2=-dt^2+a^2\,\delta_{ij} dx^i dx^j \,,
\label{FRWMetric}
\end{equation}
where $a$ is the scale factor which is a function of the cosmic time $t$ only. 
Hence, the background equations are given by
\bea
\label{00}
  f - 2 T f_{,T} - f_1 \rho_m +\rho_m f_{1, X} \dot{\phi}^2+\rho_m f_{1, Z} \dot{\phi} && \nonumber\\
  +f_2-f_{2, X} \dot{\phi}^2-f_{2, Z} \dot{\phi} =\rho_m+ \rho_r,&& \ \ \ \ \ \nonumber\\
  && \\
\label{ii}
 f - 2 T f_{,T} - 4 \dot{H} f_{,T} - 4 H \dot{f}_{,T} + f_2 =-p_r, &&\nonumber\\
  && \\
\label{phi}
 (-\rho _m f_{1,ZZ}-\rho _m f_{1,X} -2 \dot{\phi } \rho _m f_{1, X Z}-2 X \rho _m f_{1, XX}&& \nonumber\\
 +f_{2, ZZ}+f_{2, X} +2 \dot{\phi } f_{2, X Z}+2 X f_{2, XX}) \ddot{\phi} && \nonumber\\
  +3 H (\dot{\phi}f_{2,X}+f_{2, Z}) + \rho _m f_{1,\phi}-f_{2,\phi} \dot{\phi } f_{2, \phi Z}-\dot{\phi } \rho _m f_{1, \phi Z} && \nonumber\\
  2 X f_{2, \phi X}-2 X \rho _m f_{1, \phi X}- f_{,\phi}=0, && \nonumber\\
  && 
\eea
where $H\equiv \dot{a}/a$ is the Hubble rate, a dot represents derivative with respect to $t$, and a comma denotes derivative with respect to $\phi$, $X$, $Z$ or $T$.

Then, by using the following definitions for the arbitrary functions $f$, $f_1$ and $f_2$ \cite{amendola2020scaling}:
\bea
&&f=-\frac{1}{2\kappa^2}T-\frac{F(\phi)T}{6},
\label{eq:9}\\
&&f_1=\frac{1}{V_2(\phi)}-1,
\label{eq:10}\\
&&f_2=X\left[1-\frac{1}{Y_1}+2^{1-s/2}\beta\left(\frac{{Y_2}^s}{{Y_1}^{s/2}}\right)\right],
\label{eq:11}
\eea

where $s$ is a constant \cite{amendola2020scaling}, $Y_1=X/V_1(\phi)$ and $Y_2=Z/\sqrt{V_1(\phi)}$, the background equations \eqref{00}-\eqref{phi} become
\bea
\frac{3 H^2}{\kappa ^2}=&&\ -F H^{2}+\beta  \dot{\phi}^2+\rho_r+V_1+\tilde{\rho}_m+\frac{1}{2} \dot{\phi}^2,\label{eq:12}\\
-\frac{2 \dot{H}}{\kappa ^2}=&&\ \frac{2}{3} H \dot{\phi} F_{,\phi}+\frac{2}{3} F \dot{H}+2 \beta  \dot{\phi}^2+\frac{4}{3}\rho_r+\tilde{\rho}_m+\dot{\phi}^2, \label{eq:13} \nonumber\\
  && 
\eea
\bea
(1+2\beta)\ddot{\phi}+H^{2} F_{,\phi}+3H \dot{\phi}(2\beta +1)+{V_1}_{,\phi}-\frac{\tilde{\rho}_m {V_2}_{,\phi}}{{V_2}}=0,\label{eq:14} \nonumber\\
  && 
\eea
where we have defined $\tilde{\rho}_m\equiv \rho_m/V_2$ as an effective matter energy density variable.

Following Ref. \cite{Copeland:2006wr} one can rewrite the Friedmann equations \eqref{eq:12} and \eqref{eq:13} in their standard form as
\bea
\label{SH00}
&& \frac{3}{\kappa^2} H^2=\rho_{de}+\tilde{\rho}_{m}+\rho_{r},\\
&& -\frac{2}{\kappa^2} \dot{H}=\rho_{de}+p_{de}+\tilde{\rho}_{m}+\frac{4}{3}\rho_{r},
\label{SHii}
\eea where the effective energy and pressure densities are defined as 
{\small
\bea
\label{rhode}
\rho_{de}&=&\left(\beta+\frac{1}{2}\right)\dot{\phi}^2+V_1-F H^{2} ,\\[10pt]
 p_{de}&=& \frac{2}{3}H \dot{\phi} F_{,\phi}+\frac{2}{3} F \dot{H}+F H^{2}+\left(\beta+\frac{1}{2}\right)\dot{\phi}^2-V_1.
 \label{pde}
\eea }
Then, the effective dark energy equation-of-state (EOS) parameter is
\begin{equation}
w_{de}=\frac{p_{de}}{\rho _{de}}.
\label{wDE1}
\end{equation}
For these definitions of $\rho_{de}$ and $p_{de}$ one can verify that they satisfy
\begin{eqnarray}
\dot{\rho}_{de}+3H(\rho_{de}+p_{de})=-\rho_m \dot{f_1}.
\end{eqnarray} This equation is consistent with the energy conservation law and the fluid evolution equations
\bea
\label{rho_m}
&& \dot{\tilde{\rho}}_{m}+3 H\tilde{\rho}_{m}=+\rho_m \dot{f_1},\\
&& \dot{\rho}_{r}+4 H\rho_{r}=0.
\label{rho_r}
\eea

It is also useful to introduce the total equation of state (EoS) parameter as
\begin{equation}
w_{\text{tot}} = \frac{p_{\text{de}} + p_r}{\rho_{\text{de}} + \tilde{\rho}_{m} + \rho_r},
\label{wtot}
\end{equation}
which is related to the deceleration parameter \( q \) through
\begin{equation}
q = \frac{1}{2} \left(1 + 3w_{\text{tot}}\right).
\label{deccelparam}
\end{equation}
Thus, the Universe undergoes acceleration for \( q < 0 \), or equivalently for \( w_{\text{tot}} < -\frac{1}{3} \).

Finally, another useful set of cosmological parameters we can introduce is the standard density parameters:
\begin{equation}
\Omega_{m} \equiv \frac{\kappa^2 \tilde{\rho}_m}{3 H^2}, \quad \Omega_{\text{de}} \equiv \frac{\kappa^2 \rho_{\text{de}}}{3 H^2}, \quad \Omega_r \equiv \frac{\kappa^2 \rho_r}{3 H^2},
\end{equation}
which satisfies the constraint equation
\begin{equation}
\Omega_{\text{de}} + \Omega_m + \Omega_r = 1.
\end{equation}

This equation constrains the energy density of each component of the Universe in the same way as the Friedmann equation \eqref{SH00}, but expressed in terms of the density parameters.

A detailed dynamical analysis of this model is performed in the following section, where we construct the corresponding dynamical system from equations \eqref{SH00}, \eqref{SHii}, \eqref{rho_m}, and \eqref{rho_r}.


\subsection{Phase space Analysis}\label{phase_space}
In order to obtain the dynamical system of the model, we introduce the following set of dimensionless variables \cite{Copeland:2006wr}:


\begin{table*}[ht]
 \centering
 \caption{Critical points for the autonomous system.}
\begin{center}
\begin{tabular}{c c c c c c c c}\hline\hline
Name &  $x_c$ & $y_c$ & $u_{c }$   & $\varrho_{c}$ \\  \hline\\
$\ \ \ \ \ \ \ \ a_{R} \ \ \ \ \ \ \ \ $ & $0$ & $0$ & $0$   & $1$  \\ \\
$\ \ \ \ \ \ \ \ b \ \ \ \ \ \ \ \ $ & $0$ & $\sqrt{\dfrac{\sigma}{\lambda+\sigma}}$ & $\dfrac{\lambda}{\lambda+\sigma}$  & $0$   \\
$\ \ \ \ \ \ \ \ c_{R} \ \ \ \ \ \ \ \ $ & $ -\dfrac{1}{\sqrt{6}Q}$ & $0$ & $0$ & $ \ \ \ \ \ \ \ \ \frac{\sqrt{Q^2-\beta-\frac{
1}{2}}}{Q} \ \ \ \ \ \ \ \ $   \\[15pt]
$\ \ \ \ \ \ \ \ d_M \ \ \ \ \ \ \ \ $ & $-\sqrt{\dfrac{2}{3}}\dfrac{Q}{1+2\beta}$ & $0$ & $0$ & $0$  \\[15pt]
$\ \ \ \ \ \ \ \ e^{\pm} \ \ \ \ \ \ \ \ $ & $\pm \dfrac{1}{\sqrt{1+2\beta}}$ & $0$ & $0$ & 0    \\[15pt]
$\ \ \ \ \ \ \ \ f_{R} \ \ \ \ \ \ \ \ $ & $\sqrt{\dfrac{2}{3}}\dfrac{2}{\lambda}$ & $\sqrt{\dfrac{1+2\beta}{3}}\dfrac{2}{\lambda}$ & 0  &$\dfrac{\sqrt{\lambda^2-8\beta-4}}{\lambda}$\\[15pt]
$\ \ \ \ \ \ \ \ g \ \ \ \ \ \ \ \ $ & $\dfrac{\lambda}{\sqrt{6}(1+2\beta)}$ & $\sqrt{\dfrac{6+12\beta-\lambda^2}{6(1+2\beta)}}$ & 0  &$0$\\[15pt]
$\ \ \ \ \ \ \ \ h \ \ \ \ \ \ \ \ $ & $\sqrt{\dfrac{3}{2}}\dfrac{1}{Q+\lambda}$ & $ \ \ \ \ \ \ \ \ \dfrac{\sqrt{3+2Q^2+6\beta+2Q\lambda}}{\sqrt{2}(Q+\lambda)} \ \ \ \ \ \ \ \ $ & 0  &$0$ \\[15pt]
$\ \ \ \ \ \ \ \ i_M \ \ \ \ \ \ \ \ $ & $0$ & $0$ & $\dfrac{Q}{Q-\sigma }$  &$0$ \\[15pt]

\hline\hline
\end{tabular}
\end{center}
\label{table1}
\end{table*}
\begin{table*}[ht]
 \centering
 \caption{Cosmological parameters for the critical points shown in Table \ref{table1}}
\begin{center}
\begin{tabular}{c c c c c c c c c}\hline\hline
Name &  $\Omega_{de}$ & $\Omega_{m}$ & $\Omega_{r}$ & $w_{de}$ & $w_{tot}$\\  \hline\\
$\ \ \ \ \ \ \ \ a_{R} \ \ \ \ \ \ \ \ $ & $0$ & $0$ & 1 & $\dfrac{1}{3}$ & $\dfrac{1}{3}$ \\[15pt]
$\ \ \ \ \ \ \ \ b \ \ \ \ \ \ \ \ $ &  1 & 0 & 0 & $-1$ & $-1$ \\[15pt]
$\ \ \ \ \ \ \ \ c_{R} \ \ \ \ \ \ \ \ $ &  $\dfrac{1+2\beta}{6Q^2}$ & $\dfrac{1+2\beta}{3Q^2}$ & $1-\dfrac{1+2\beta}{2Q^2}$ & 1 & $\dfrac{1}{3}$  \\[15pt]
$\ \ \ \ \ \ \ \ d_M \ \ \ \ \ \ \ \ $ &  $\dfrac{2 Q^2}{6 \beta +3}$ & $\dfrac{6 \beta -2 Q^2+3}{6 \beta +3}$ & 0 & 1 & $\dfrac{2 Q^2}{6 \beta +3}$ \\[15pt]
$\ \ \ \ \ \ \ \ e^{\pm} \ \ \ \ \ \ \ \ $ &  1 & 0 & 0 & 1 & 1  \\[15pt]
$\ \ \ \ \ \ \ \ f_{R} \ \ \ \ \ \ \ \ $ & $\dfrac{8 \beta +4}{\lambda ^2}$ & 0 & $1-\dfrac{8 \beta +4}{\lambda ^2}$ & $\dfrac{1}{3}$ & $\dfrac{1}{3}$ \\[15pt]
$\ \ \ \ \ \ \ \ g \ \ \ \ \ \ \ \ $ &  1 & 0 & 0 & $\dfrac{\lambda ^2}{6 \beta +3}-1$ & $\dfrac{\lambda ^2}{6 \beta +3}-1$\\[15pt]
$\ \ \ \ \ \ \ \ h \ \ \ \ \ \ \ \ $ &  $\dfrac{6 \beta +Q^2+\lambda  Q+3}{(\lambda +Q)^2}$ & $\ \ \ \ \ \ \ \ \dfrac{-6 \beta +\lambda ^2+\lambda  Q-3}{(\lambda +Q)^2}\ \ \ \ \ \ \ \ $ & 0 & $-\dfrac{Q (\lambda +Q)}{6 \beta +Q (\lambda +Q)+3}$ & $-\dfrac{Q}{\lambda +Q}$\\[15pt]
$\ \ \ \ \ \ \ \ i_M \ \ \ \ \ \ \ \ $ &  $\dfrac{Q}{Q-\sigma }$ & $1-\dfrac{Q}{Q-\sigma }$ & 0 & 0 & 0\\[15pt]

 \hline\hline
\end{tabular}
\end{center}
\label{table2}
\end{table*}

\begin{eqnarray}
x =& \dfrac{\kappa  \dot{\phi}}{\sqrt{6} H}, \ \ \ \ \ \ y =& \dfrac{\kappa  \sqrt{V_1}}{\sqrt{3} H}, \ \ \ \ \ \ \ u = -\frac{1}{3}\kappa ^2 F,\nonumber\\ 
\Omega_m =& \dfrac{\kappa^2\rho_m }{V_2\ 3H^2}, \ \ \ \ \ \
\lambda =& - \dfrac{{V_1}_{,\phi}}{\kappa V_1}, \ \ \ \ \ \ \ \sigma = - \dfrac{F_{,\phi}}{\kappa F}, \nonumber\\
Q =&  -\dfrac{V_{2,\phi}}{\kappa V_2}, \ \ \ \ \ \ \ \Theta =&  \dfrac{F F_{,\phi\phi}}{(F_{,\phi})^2}, \ \ \ \ \ \ \ \Gamma_1 = \dfrac{V_1 V_{1,\phi\phi}}{(V_{1,\phi})^2}, \nonumber\\
 \Gamma_2 =& \dfrac{V_2 V_{2,\phi\phi}}{(V_{2,\phi})^2},\ \ \ \ \ \ \ \varrho= &\frac{\kappa\sqrt{\rho_r}}{\sqrt{3}H},\ \nonumber\\
&& 
\label{var}
\end{eqnarray}
and the constraint equation
\begin{equation}
    \Omega_m + \varrho^2 +u+(1+2\beta) x^2+y^2= 1.
\end{equation}
Therefore, we obtain the dynamical system
\bea
\dfrac{\text{d}x}{\text{d}N}=&&\ F_1 (x,y,\varrho, u, \lambda, Q, \sigma), \label{dysys1}\\
\dfrac{\text{d}y}{\text{d}N}=&&\ \frac{1}{2 (u-1)} F_2 (x,y,\varrho, u, \lambda, Q, \sigma),\\
\dfrac{\text{d}\varrho}{\text{d}N}=&&\ -\frac{\varrho  F_3 (x,y,\varrho, u, \lambda, Q, \sigma)}{2 (u-1)},\\
\dfrac{\text{d}u}{\text{d}N}=&&-\sqrt{6} \sigma  x u,  \label{dysys2}\\
\dfrac{\text{d}\lambda}{\text{d}N}=&&-\sqrt{6} (\Gamma_1-1) \lambda ^2 x,\\
\dfrac{\text{d}Q}{\text{d}N}=&&-\sqrt{6} (\Gamma_2-1) Q^2 x,\\
\dfrac{\text{d}\sigma}{\text{d}N}=&&-\sqrt{6} (\Theta -1) \sigma ^2 x, \label{dysys3}
\label{ODE10}
\eea
where,
\bea
F_1 &=& \frac{1}{2} \left[\frac{\sqrt{6} \left(Q \left((2 \beta +1) x^2+y^2+u+\varrho ^2-1\right)+\lambda  y^2-\sigma  u\right)}{2 \beta +1}\right.\nonumber\\ 
&-&\left.\frac{3 (2 \beta +1) x^3-2 \sqrt{6} \sigma  x^2 u-x \left(-3 y^2+3 u+\varrho ^2-3\right)}{u-1}\right],\nonumber \\
F_2 &=& y \Bigg[\sqrt{6} x (\lambda -\lambda u-2 \sigma  u)+3 u-3 (2 \beta +1) x^2+ \nonumber\\ 
&&+3 y^2-\varrho ^2-3\Bigg], \nonumber\\
F_3 &=& (6 \beta +3) x^2+2 \sqrt{6} \sigma  x u-3 y^2+u+\varrho ^2-1. \nonumber
\eea

Using the above set of phase space variables, we can also write
\bea
\Omega_{de} &=& (1+2 \beta)x^2+y^2+u,\\[10pt]
\Omega_{m} &=& 1-(1+2 \beta)x^2-y^2-u-\varrho^2,\\[10pt]
\Omega_{r} &=& \varrho^2.
\eea

Similarly, the equation of state of dark energy $w_{de}=p_{de}/\rho_{de}$ can be rewritten as

\bea
w_{de} &=& -\frac{2 \sqrt{6} \sigma  u x+u \varrho ^2+(6 \beta +3) x^2-3 y^2}{3 (u-1) \left[u+(2 \beta +1) x^2+y^2\right]},
\eea
whereas the total equation of state becomes
\bea
w_{tot}&=& \frac{1}{3} \left[\varrho ^2-\frac{2 \sqrt{6} \sigma  u x+u \varrho ^2+(6 \beta +3) x^2-3 y^2}{u-1}\right]. \nonumber\\
&&
\eea

For the dynamical system described by \eqref{dysys1}-\eqref{dysys2} to be autonomous, the parameters $\Gamma_1$, $\Gamma_2$, and $\Theta$ must be known. Therefore, we select the following exponential functions:
\begin{equation}
    V_1 \sim e^{- \kappa \lambda \phi}, \ \ \ V_2 \sim e^{- \kappa Q \phi}, \ \ \ \text{and} \ \ \ F \sim e^{- \kappa \sigma \phi}.
\end{equation}
 In this way we set the values $\Gamma_1=\Gamma_2=\Theta=1$ and establish $\lambda$, $Q$ and $\sigma$ as dimensionless constants for the model. These types of functions have been shown to lead to accelerated expansion and the derivation of scaling solutions in previous works \cite{Copeland:2006wr,amendola2010dark,amendola2020scaling,Gonzalez-Espinoza:2021qnv}.

\subsection{Critical points}

In this section, we obtain the critical points from the conditions $\text{d}{x}/\text{d}{N}=\text{d}{y}/\text{d}{N}=\text{d}{\varrho}/\text{d}{N}=\text{d}{u}/\text{d}{N}=0$, considering $V_1  \sim e^{- \kappa \lambda \phi}$, $V_2  \sim e^{- \kappa Q \phi}$ and $F  \sim e^{- \kappa \sigma \phi}$ \cite{Copeland:2006wr,Amendola:2020ldb}. Where we consider the definition of each dynamical variable \eqref{var} and that the physically allowable critical points are given by $y_c\geq0,\ \varrho_c\geq0$ and $u_c\geq0$. The critical points of the system \eqref{dysys1}-\eqref{dysys2} are shown in Table \ref{table1} and the values of their cosmological parameters in Table \ref{table2}. Also, in this subsection and further on, we introduce the parameters $\Omega_{de}^{(r)}$ and $\Omega_{de}^{(m)}$, representing the fractional density of dark energy during the radiation-dominated and dark matter-dominated eras, respectively. The conditions of existence and acceleration associated with the parameters of each critical point are presented in Table \ref{tab:A1}.\\

The critical point $a_R$ corresponds to a radiation with $\Omega_r=1$ and $w_{de}=w_{tot}=1/3$. Point $b$ is a de Sitter solution with $\Omega_{de}=1$, and $w_{de}=w_{tot}=-1$, which provides accelerated expansion for all values of the parameters. Critical point $c_R$ represents a scaling radiation era, for which $\Omega^{(r)}_{de}=(1+2\beta)/6Q^2$, $w_{de}=1$,  and $w_{tot}=1/3$. It should satisfy the early constraint imposed by the physics of Big Bang Nucleosynthesis (BBN), ensuring $\Omega^{(r)}_{de}<0.045$ \cite{Ferreira:1997hj,Bean:2001wt}. 

On the other hand, for $Q=0$, the critical point $d_M$ represents a matter-dominated era with $\Omega_m = 1$, $w_{de}=1,$
and $w_{tot} = 0$. For $Q\neq 0 $, we have a scaling matter era with $\Omega_{de}=2 Q^2/(6 \beta +3)$, which is constrained to satisfy  $\Omega_{de}^{(m)}< 0.02$ ($95\%$ C.L.), at redshift $z\approx 50$, according to CMB measurements \cite{Ade:2015rim}. In both cases of $Q$, dark energy behaves like stiff matter with $w_{de}=1$. Since $w_{tot}=2 Q^2/(6 \beta +3)$, this point presents acceleration for $-(1+2Q^2)/2<\beta<-1/2$. 

Point $e^{\pm}$ is a dark energy dominated solution which satisfies $\Omega_{de}=1$, but it cannot explain the current accelerated expansion due to its behavior as stiff matter with $w_{de} = w_{tot} = 1$.\\

The critical point $f_R$ corresponds to a scaling radiation era, where $\Omega^{(r)}_{de}=1-(8\beta+4)/\lambda^2$. Therefore, it should satisfy the early constraint imposed by the physics of BBN, ensuring $\Omega^{(r)}_{de}<0.045$ \cite{Ferreira:1997hj,Bean:2001wt}. It is worth noting that dark energy behaves as a radiation fluid with $w_{de} = w_{tot} = 1/3$.\\

Point $g$ provides a dark energy-dominated era, which can explain the cosmic accelerated expansion when $w_{tot}<-1/3$.

The point labeled as $h$ represents a matter-scaling era. As shown in Table \ref{table2}, for this fixed point, the values of the cosmological parameters depend on both the energy and momentum couplings. This point is constrained to satisfy  $\Omega_{de}^{(m)}< 0.02$ ($95\%$ C.L.), at redshift de $z\approx 50$, according to CMB measurements \cite{Ade:2015rim}. In the case where $Q=0$ satisfies $w_{de} = w_{tot} = 0$, behaving as cold dark matter. For $Q\neq0$, this fixed point can provide accelerated expansion and domination of dark energy over matter. 


Finally, point $i_M$ represents a matter-scaling era, characterized by $\Omega_{de}=Q/(Q-\sigma)$, $\Omega_m = 1-Q/(Q-\sigma)$, and $w_{de} = w_{tot} = 0$. As noted, these values depend on both the energy coupling $Q$ and the non-minimal coupling to gravity $\sigma$. 

In the following section, we will examine the stability conditions for these critical points. These conditions are determined through a linear analysis of perturbations in the phase space variables.


\subsection{Stability}

To determine the stability of critical points, we perturb the autonomous system \eqref{dysys1}-\eqref{ODE10} using linear perturbations $\delta x$, $\delta y$, $\delta \varrho$, and $\delta u$. The stability of each critical point is determined by examining the sign of the eigenvalues $\mu$ of the four-dimensional Jacobian matrix evaluated at each critical point \cite{Copeland:2006wr}. 
\begin{itemize}
    \item Point $a_R$ has the eigenvalues
    \bea
    \mu_1=-1\ , \ \mu_2=1\ , \ \mu_3=0\ , \ \mu_4=2. 
    \eea

    \item Point $b$ has the eigenvalues
    \bea
    \mu_1&=&-3\ , \ \mu_2=-2\ , \nonumber\\[7pt] 
    \hspace{4mm}\mu_{3,4}&=&-\frac{3+6\beta\pm\sqrt{3(1+2\beta)(3+6\beta+4\lambda\sigma)}}{2+4\beta}.\nonumber\\
    &&
    \eea

    \item Point $c_R$ has the eigenvalues
    \bea
    \mu_{1,2}&=&\frac{1}{2}\left[-1\mp\sqrt{\frac{2-3Q^2+4\beta}{Q^2}}\right]\ , \ \mu_3=2+\frac{\lambda}{2Q}\ , \nonumber\\[7pt] 
    \mu_{4}&=&\frac{\sigma}{Q}.
    \eea

    \item Point $d_M$ has the eigenvalues
    \bea
    \mu_{1}&=&-\frac{1}{2}+\frac{Q^2}{1+2\beta}\ , \ \mu_2=-\frac{3}{2}+\frac{Q^2}{1+2\beta}\ , \nonumber\\[7pt] 
    \mu_{3}&=&\frac{2Q\sigma}{1+2\beta}\ , \ \mu_4=\frac{3+6\beta+2Q(Q+\lambda)}{2+4\beta}.
    \eea

    \item Point $e^{\pm}$ has the eigenvalues
    \bea
    \mu_{1}&=&1\ , \ \mu_2=3+\frac{\sqrt{6}Q}{\sqrt{1+2\beta}}\ , \nonumber\\[7pt] 
    \mu_{3}&=&3-\frac{\sqrt{3}\lambda}{\sqrt{2(1+2\beta)}}\ , \ \mu_4=-\frac{\sqrt{6}\sigma}{\sqrt{1+2\beta}}.
    \eea

    \item Point $f_R$ has the eigenvalues
    \bea
    \mu_1 &=& 1+\frac{4Q}{\lambda}\ , \ \mu_2=-\frac{4\sigma}{\lambda}\ , \nonumber\\[7pt] 
    \mu_{3,4} &=& -\frac{\lambda\pm\sqrt{64+128\beta-15\lambda^2}}{2\lambda}.
    \eea

    \item Point $g$ has the eigenvalues
    \bea
    \mu_{1}&=&-3+\frac{\lambda^2}{2+4\beta}\ , \ \mu_2=\frac{-3-6\beta+\lambda(Q+\lambda)}{1+2\beta}\ , \nonumber\\[7pt] 
    \mu_{3}&=&-2+\frac{\lambda^2}{2+4\beta}\ , \ \mu_4=-\frac{\lambda\sigma}{1+2\beta}.
    \eea

    \item Point $h$ has the eigenvalues
    {\small
    \bea
    \mu_1&=&-\frac{4Q+\lambda}{2(Q+\lambda)}\ , \ \mu_2=-\frac{3\sigma}{Q+\lambda}\ , \nonumber\\[7pt] 
    \mu_{3,4}&=&-\frac{1}{4}\Bigg[-6+\frac{3\lambda}{Q+\lambda}\pm\nonumber\\
    && \sqrt{-63-\frac{48Q\lambda}{1+2\beta}+\frac{9(24+Q^2+48\beta)}{(Q+\lambda)^2}+\frac{234Q}{Q+\lambda}}\Bigg].
    \eea}

    \item Point $i_M$ has the eigenvalues
    \bea
    \mu_1&=&-\dfrac{1}{2}\ , \ \mu_2=\dfrac{3}{2}\ , \nonumber\\[7pt] 
    \mu_{3,4}&=&\frac{1}{4}\left[-3\mp\sqrt{9-\frac{48Q\sigma}{1+2\beta}}\right].
    \eea
\end{itemize}

Table \ref{tab:A1} presents a detailed description of the stability conditions for each critical point, its corresponding eigenvalues, and the parameter constraints. 

\begin{figure}[!h]
    \centering
        \includegraphics[scale=0.25]{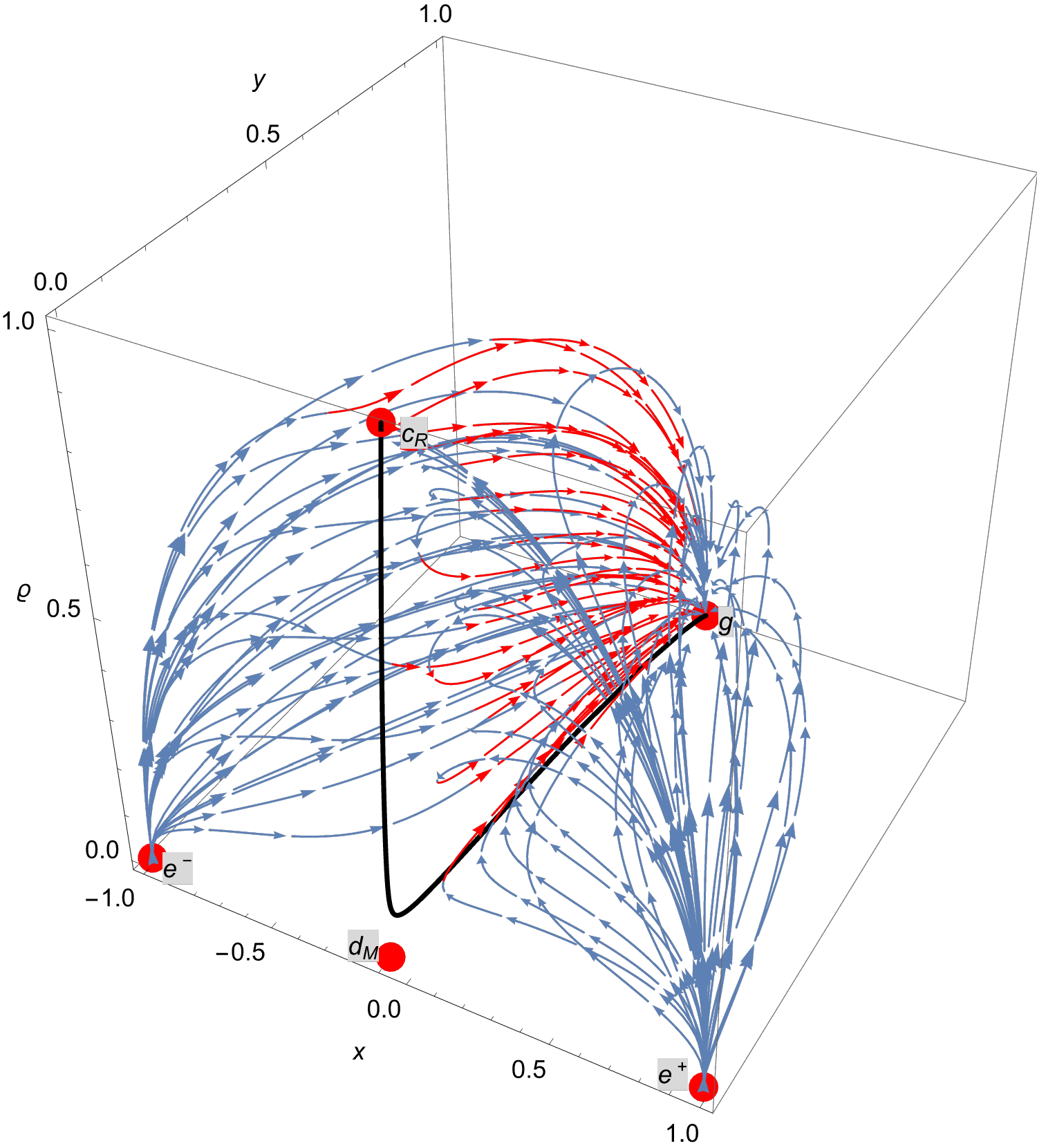}
        \caption{Phase space stream plot for the values $\beta = 0.01$, $Q = 0.1$, $\lambda = 0.1$ and $\sigma = 0.1$. The black-solid curve corresponds to the evolution curve and represents the physical trajectory of the three-dimensional system with initial conditions $x_0 = 10^{-11}$, $y_0 = 7.4 \times 10^{-13}$, $u_0 = 10^{-12}$ and $\varrho = 0.99983$.}  
    \label{3Dphase1}
\end{figure}

\begin{figure}[!h]
    \centering
        \includegraphics[scale=0.25]{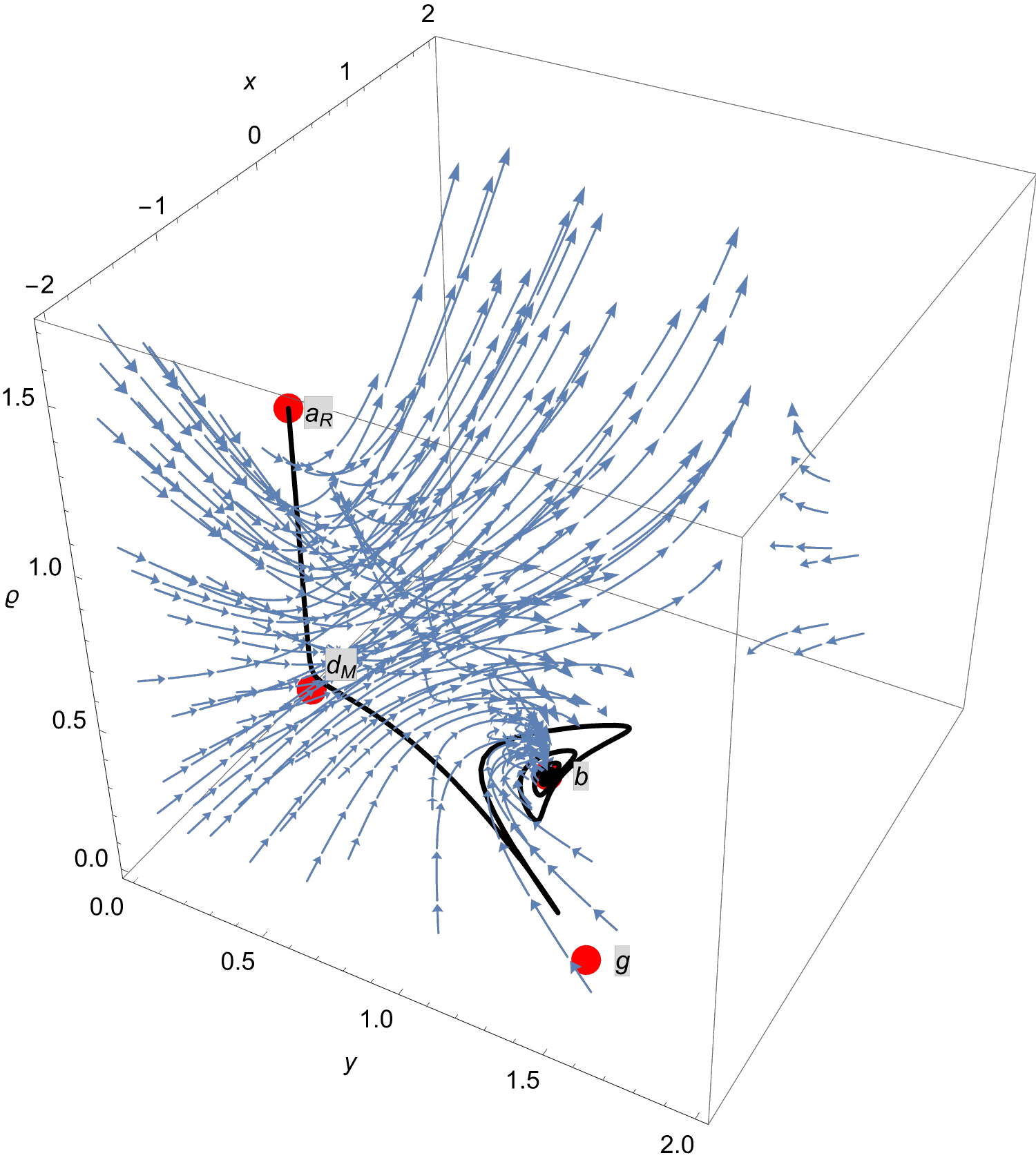}
        \caption{Phase space stream plot for the values $\beta = -0.8$, $Q =  -4.0 \times 10^{-3}$, $\lambda = 2$ and $\sigma = 17$. The black-solid curve corresponds to the evolution curve and represents the physical trajectory of the three-dimensional system with initial conditions $x_0 = 10^{-11}$, $y_0 = 4.9 \times 10^{-13}$, $u_0 = 10^{-12}$ and $\varrho = 0.99983$.}  
    \label{3Dphase2}
\end{figure}


\begin{figure}[!h]
    \centering
        \includegraphics[scale=0.25]{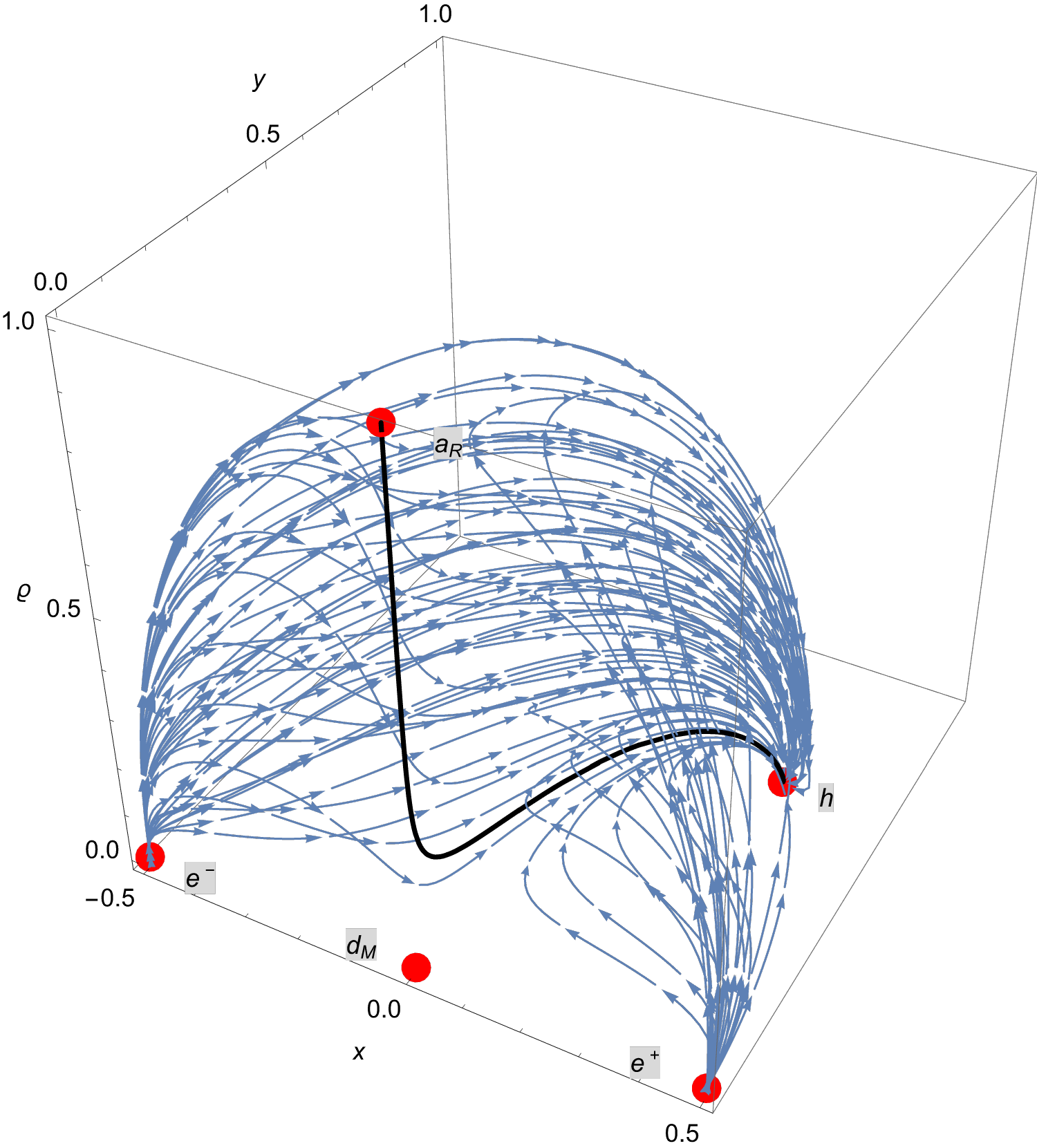}
        \caption{Phase space stream plot for the values $\beta = 1.5$, $Q =  3.0 \times 10^{-2}$, $\lambda = 3.75$ and $\sigma = 1$. The black-solid curve corresponds to the evolution curve and represents the physical trajectory of the three-dimensional system with initial conditions $x_0 = 10^{-9}$, $y_0 = 5.1 \times 10^{-11}$, $u_0 = 10^{-9}$ and $\varrho = 0.99983$.}  
    \label{3Dphase3}
\end{figure}

\begin{figure}[!h]
    \centering
        \includegraphics[scale=0.35]{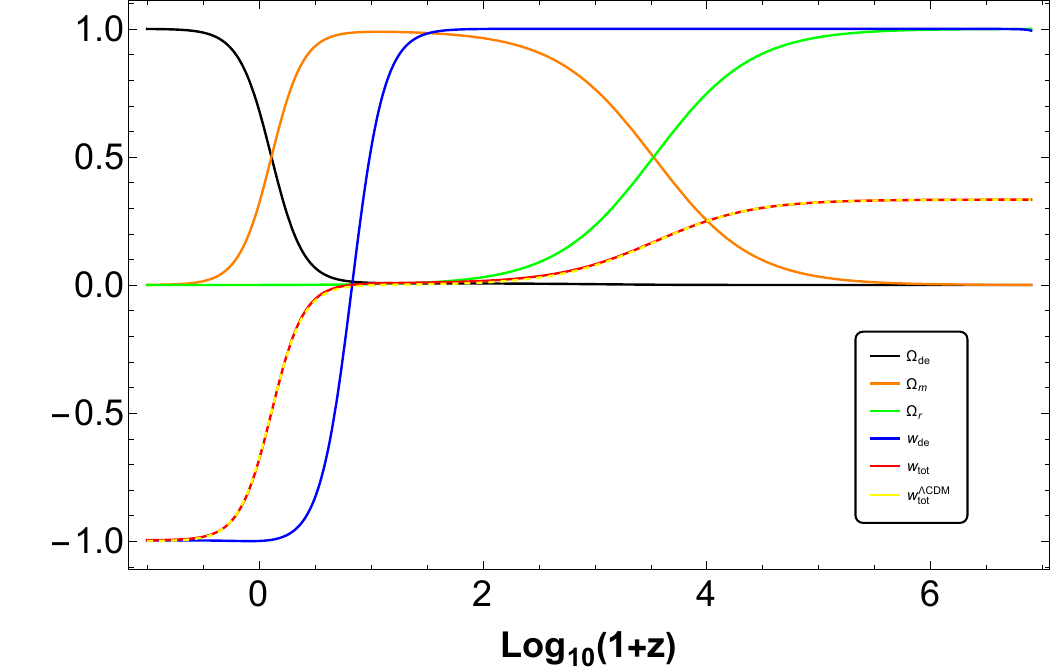}
        \caption{We depict the evolution of the fractional energy of dark energy $\Omega_{de}$ (black), dark matter (including baryons) $\Omega_m$ (orange), radiation $\Omega_r$ (green), equation of state parameter of dark energy $w_{de}$ (blue), total EoS parameter $w_{tot}$ (red) and the EoS parameter of the $\Lambda$CDM model (yellow) as functions of the cosmological redshift, for the same initial conditions used in Fig. \ref{3Dphase1}.}  
    \label{wtotal1}
\end{figure}

\begin{figure}[!h]
    \centering
        \includegraphics[scale=0.35]{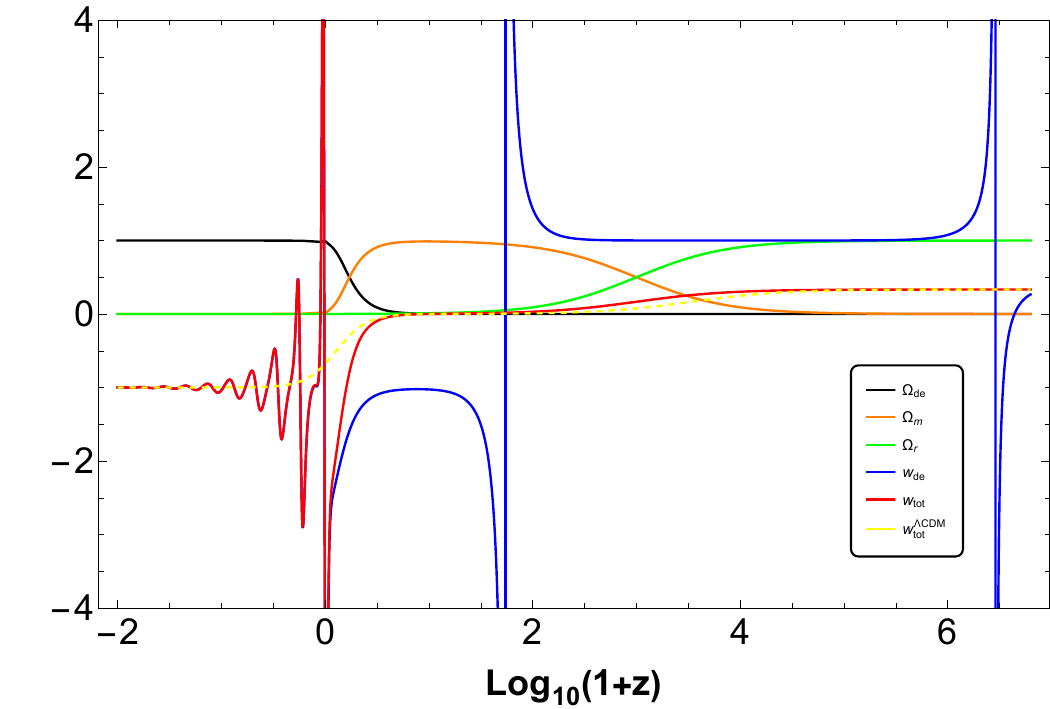}
        \caption{We depict the evolution of the fractional energy of dark energy $\Omega_{de}$ (black), dark matter (including baryons) $\Omega_m$ (orange), radiation $\Omega_r$ (green), equation of state parameter of dark energy $w_{de}$ (blue), total EoS parameter $w_{tot}$ (red) and the EoS parameter of the $\Lambda$CDM model (yellow) as functions of the cosmological redshift, for the same initial conditions used in Fig. \ref{3Dphase2}.}  
    \label{wtotal2}
\end{figure}

\begin{figure}[!h]
    \centering
        \includegraphics[scale=0.35]{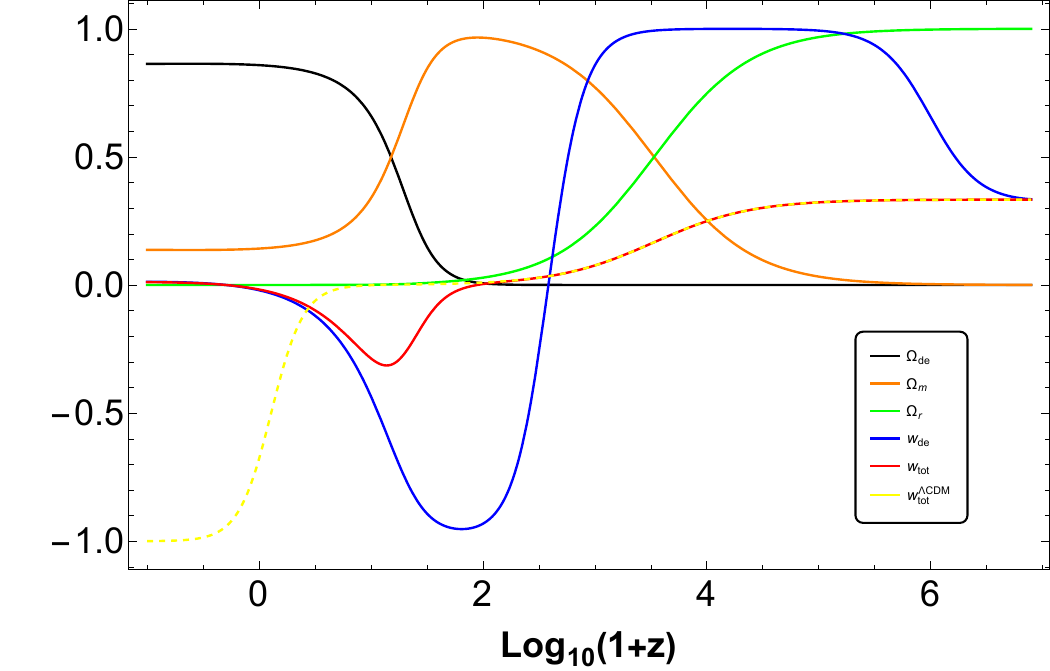}
        \caption{We depict the evolution of the fractional energy of dark energy $\Omega_{de}$ (black), dark matter (including baryons) $\Omega_m$ (orange), radiation $\Omega_r$ (green), equation of state parameter of dark energy $w_{de}$ (blue), total EoS parameter $w_{tot}$ (red) and the EoS parameter of the $\Lambda$CDM model (yellow) as functions of the cosmological redshift, for the same initial conditions used in Fig. \ref{3Dphase3}.}  
    \label{wtotal3}
\end{figure}


\begin{figure}[!h]
    \centering
        \includegraphics[scale=0.35]{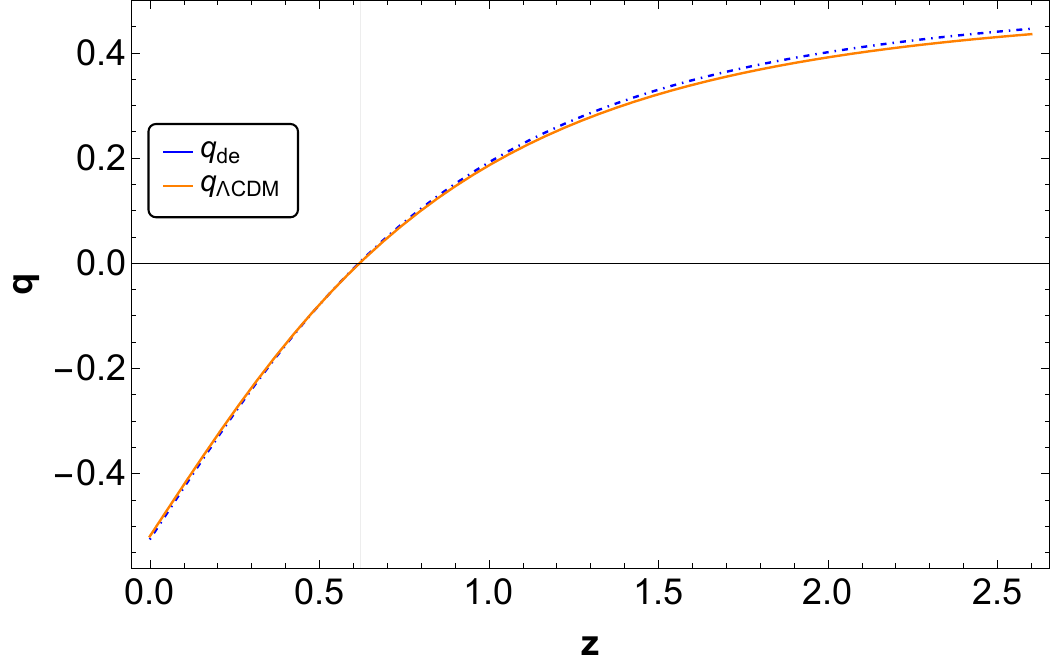}
        \caption{We present the evolution of the deceleration parameter $q(z)$, calculated using the same initial conditions as those employed in Fig. \ref{3Dphase1}. We also present the evolution curve of the deceleration parameter $q_{\Lambda\mathrm{CDM}}(z)$ of the $\Lambda$CDM model.}  
    \label{qtotal}
\end{figure}

\begin{figure}[!h]
    \centering
        \includegraphics[scale=0.35]{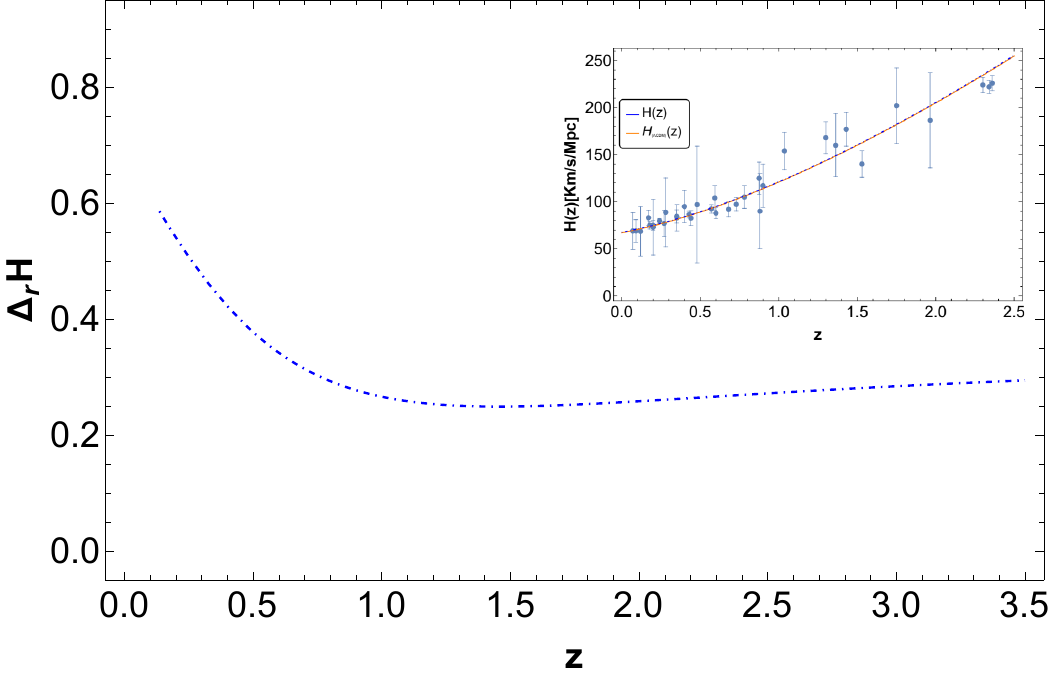}
        \caption{We present the evolution of the Hubble rate $H(z)$ and its relative difference $\Delta_r H(z) = 100\times \frac{|H-H_{\Lambda\text{CDM}}|}{H_{\Lambda\text{CDM}}}$ for the $\Lambda\text{CDM}$ model as functions of redshift, using the same initial conditions depicted in Fig. \ref{3Dphase1}. Additionally, we provide the evolution of the Hubble rate $H_{\Lambda\text{CDM}}$ within the $\Lambda$CDM framework, alongside observational Hubble data sourced from \cite{cao2018cosmological,farooq2013hubble}. The present-day Hubble rate, $H_0 = 67.4$ km/(Mpc·s), as reported by Planck 2018 \cite{Aghanim:2018eyx}, has been utilized in our analysis.}  
    \label{dHtotal}
\end{figure}

\section{Numerical analysis}\label{num_a}

In this section, we conduct a numerical analysis of the autonomous system \eqref{dysys1}-\eqref{dysys2}. Our investigation examines how well our model explains the current accelerated expansion of the Universe and compares our predictions with the latest observational data on cosmological parameters. 

Figures \ref{3Dphase1}, \ref{3Dphase2}, and \ref{3Dphase3} illustrate the phase space stream flow for the trajectories $c_R \to d_M \to g$, $a_R \to d_M \to g \to b$, and $a_R \to d_M \to h$. It is evident that the solutions of the autonomous system converge to the attractors $b$ and $g$, as well as to the new solution $h$ for specific parameter values. Although the system exhibits a scaling attractor behavior, the primary physical trajectory of interest is $c_R \to d_M \to g$, highlighted by a red stream flow in Fig. \ref{3Dphase1}.

In Fig. \ref{wtotal1}, we depict the behavior of the fractional energies of dark energy, matter (including baryons), and radiation, as well as the total EoS parameter and EoS parameter of dark energy for the physical trajectory $c_R\to d_M\to g$. Similarly, in Figs. \ref{wtotal2} and \ref{wtotal3}, we present the evolution of the fractional energy densities  and the EoS parameter for the physical trajectories $a_{R}\rightarrow d_{M}\rightarrow g\rightarrow b$ and $a_{R}\rightarrow d_{M}\rightarrow h$, respectively. The transition between the radiation and matter eras occurs around $z \approx 3387$, while the transition to the accelerated phase occurs around $z \approx 0.62$, as indicated by the deceleration parameter shown in Fig. \ref{qtotal}. These values are very close to those predicted by the $\Lambda$CDM model and are consistent with current observational data \cite{Aghanim:2018eyx}.
Also, we have obtained the fractional energy density parameters of dark energy $\Omega_{de}\approx0.68$ and matter $\Omega_{m}\approx0.32$ with the equation of state of dark energy at $z=0$ given by $w_{de}^{(0)}\approx-0.999392$, which is consistent with the observational constraint $w_{de}^{(0)}=-1.028\pm0.032$ from the latest observations \cite{Aghanim:2018eyx}. 
We have also applied the constraint on the fractional energy density of dark energy during the scaling matter regime, derived from Planck CMB measurements, $\Omega_{de}^{(m)} < 0.02\ (95\%\ \mathrm{C.L.})$ at a redshift of approximately $z \approx 50$ \cite{Ade:2015rim}. As shown in Fig. \ref{wtotal1}, we find $\Omega_{de}^{(m)} \approx 0.006$ at redshift $z = 50$ during the scaling matter era $d_M$.


Finally, in Fig. \ref{dHtotal}, we present an analysis of the Hubble rate (see Appendix \ref{appen_H}) by calculating the evolution of the Hubble rate $H(z)$ in our model, alongside the evolution of the Hubble rate $H_{\Lambda\mathrm{CDM}}(z)$ for the $\Lambda$CDM model. This comparison uses the same parameter values and initial conditions as those in Fig. \ref{3Dphase1}. The relative difference, $\Delta_r H(z)$, between our model and the $\Lambda$CDM model is also shown, demonstrating their close correspondence and consistent alignment with the latest observational data.

\section{Concluding Remarks}\label{Remarks}

We investigated the cosmological behavior of dark energy within a model that includes a scalar field non-minimally coupled to torsion gravity. This model also accounts for the interaction between the scalar field and cold dark matter through energy and momentum transfer. In this framework, torsion emerges from the Weitzenböck connection in teleparallel gravity, which is a flat connection that exhibits non-zero torsion in the presence of gravity. We derived the cosmological equations and formulated the corresponding autonomous system. A detailed phase space analysis was conducted, where we identified all critical points and determined their stability conditions. Additionally, we compared our theoretical predictions with the latest observational data from $H(z)$ measurements.


Our analysis identified a stable critical point, $g$, which corresponds to the current accelerated expansion of the Universe, resembling a quintessence-like fixed point. We also found a de Sitter attractor solution, $b$, which is a spiral stable point. This suggests that, regardless of the initial conditions—provided they are close to these attractor points—the system will naturally evolve toward a dark energy-dominated phase characterized by accelerated expansion. Additionally, we demonstrated that this late-time accelerated phase can be smoothly connected to the standard radiation and matter-dominated eras.


We also identified the presence of a matter-scaling era ($d_M$) associated with the interaction between dark energy and dark matter through the exchange of energy and momentum. Furthermore, we discovered a second fixed point, also a matter-scaling era ($i_M$), which arises due to the energy transfer from the scalar field to dark matter and its non-minimal coupling to gravity. Consequently, we determined the necessary conditions for the model parameters that allow the system to transition from these scaling regimes to a dark energy-dominated attractor with acceleration.


Finally, we obtained a scaling attractor solution ($h)$ in which the energy density of the universe is dominated by the field density. These types of solutions are particularly interesting because they offer a natural mechanism for alleviating the coincidence problem \cite{Copeland:2006wr}. However, in the context of the present model, this attractor solution is unable to simultaneously account for both the current accelerated expansion and the thermal history of the universe.

\begin{acknowledgments}
M. Gonzalez-Espinoza acknowledges the financial support of FONDECYT de Postdoctorado, N° 3230801. C. Rodriguez-Benites and M. Alva-Morales acknowledge the financial support of PE501082885-2023-PROCIENCIA. G. Otalora gratefully acknowledges the hospitality of the {\it Institute of Cosmology and Gravitation (ICG)} at the University of Portsmouth, where part of this work was carried out.

\end{acknowledgments}

\hfill \break

\newpage

\bibliography{bio,bio2} 

\hfill \break
\hfill \break
\hfill \break
\hfill \break
\hfill \break
\hfill \break
\hfill \break
\hfill \break
\hfill \break
\hfill \break
\hfill \break
\hfill \break
\hfill \break
\hfill \break
\hfill \break
\hfill \break
\hfill \break
\hfill \break
\hfill \break
\hfill \break
\hfill \break
\hfill \break
\hfill \break
\hfill \break
\hfill \break
\hfill \break
\hfill \break
\hfill \break
\hfill \break
\hfill \break
\hfill \break
\hfill \break
\hfill \break
\hfill \break
\hfill \break
\hfill \break
\hfill \break
\hfill \break
\hfill \break
\hfill \break
\hfill \break
\hfill \break
\hfill \break
\hfill \break
\hfill \break
\hfill \break
\hfill \break
\hfill \break
\hfill \break

\newpage

\begin{appendix}

\section{Properties of critical points}

The properties of the critical points of the autonomous system \eqref{ODE10} are shown in Table \ref{tab:A1}, where the following set of definitions has been made:

\begin{itemize}
    \item For point $c_R$
    \begin{align*}
        \beta_{1,c}=\frac{1}{4}(-2+3Q^2) \ , \ \beta_{2,c}=\frac{1}{4}(-1+2Q^2).
    \end{align*}
    
    \item For point $d_M$
    \begin{align*}
        \beta_{d,1}=\frac{1}{2}(-1+2Q^2) \ , \ \beta_{d,2}=\frac{1}{2}(-1-2Q^2) \ , \\[5pt]
        \lambda_{d,1}=\frac{-4-6\beta}{\sqrt{2}} \ , \ \lambda_{d,2}=\frac{-3-2Q^2-6\beta}{2Q}.
    \end{align*}

    \item For point $e^{\pm}$
    \begin{align*}
        \beta_{1,e}=\frac{1}{12}(-6+\lambda^2) \ , \ \beta_{2,e}=\frac{1}{6}(-3+2Q^2).
    \end{align*}

    \item For point $f_R$
    \begin{align*}
        \beta_{f,1}=\frac{1}{128}(-64+15\lambda^2) \ , \ \beta_{f,2}=\frac{1}{8}(-4+\lambda^2).
    \end{align*}

    \item For point $g$
    \begin{align*}
        \beta_{1,g}=\frac{1}{12}&(-6+\lambda^2) \ , \ \beta_{2,g}=\frac{1}{4}(-2+\lambda^2)\ , \\[5pt]
        &Q_{1,g}=\frac{3+6\beta-\lambda^2}{\lambda}.
    \end{align*}

    \item For point $h$
    \begin{align*}
        &\beta_{h,0}=\frac{1}{6}(-3-2Q^2-2Q\lambda) \ , \\[5pt]
        \beta_{h,1}&=\frac{1}{96}(-48-20Q^2-12Q\lambda+7\lambda^2)-\frac{R_f}{96} \ , \\[5pt]
        &\beta_{h,2}=\frac{1}{6}(-3-2Q^2-2Q\lambda)-\frac{R_f}{96} \ , \\[5pt]
        \beta_{h,3}&=\frac{1}{96}(-48-20Q^2-12Q\lambda+7\lambda^2)+\frac{R_f}{96} \ , \\[5pt]
        &\beta_{h,4}=\frac{1}{6}(-3+Q\lambda+\lambda^2).
    \end{align*}

    where \\
    $R_f=\sqrt{400Q^4+992Q^3\lambda+888Q^2\lambda^2+344Q\lambda^3+49\lambda^4}$.
\end{itemize}

\begin{table*}[t!]
\scriptsize
 \centering
 \caption{Properties of the critical points}
\begin{center}
\begin{tabular}{c c c c c c c }\hline\hline
Name &  Existence & Stability & Acceleration   \\  \hline\\
$\ \ \ \ \ \ \ \ a_{R} \ \ \ \ \ \ \ \ $ & $\forall\ \lambda,\sigma,Q,\beta$ & unstable $\forall\ \lambda,\sigma,Q,\beta$ & never    \\ \\
$\ \ \ \ \ \ \ \ b \ \ \ \ \ \ \ \ $ & $(\sigma <0\land (\lambda <0\lor 0<\lambda <-\sigma ))$ & $(\sigma <0\land (\left(\beta <-\frac{1}{2}\land \frac{-6 \beta -3}{4 \sigma }\leq \lambda <0\right)$ & always     \\
 & $\lor (\sigma >0\land (-\sigma <\lambda <0\lor \lambda >0))$ & $\lor \left(\beta >-\frac{1}{2}\land 0<\lambda \leq \frac{-6 \beta -3}{4 \sigma }\right)))$ & \\
 & & $\lor (\sigma >0\land (\left(\beta <-\frac{1}{2}\land 0<\lambda \leq \frac{-6 \beta -3}{4 \sigma }\right)$ & \\
 & & $\lor \left(\beta >-\frac{1}{2}\land \frac{-6 \beta -3}{4 \sigma }\leq \lambda <0)\right))$ & \\[15pt]
$\ \ \ \ \ \ \ \ c_{R} \ \ \ \ \ \ \ \ $ & $\left(Q<0\land -\frac{1}{2}<\beta <\frac{1}{2} \left(2 Q^2-1\right)\right)$ & $(\lambda <0\land 0<Q<-\frac{\lambda }{4}\land \beta _{1,c}\leq \beta <\beta _{2,c}$ & never &    \\
 & $\lor \left(Q>0\land -\frac{1}{2}<\beta <\frac{1}{2} \left(2 Q^2-1\right)\right)$ & $ \land \sigma <0)\lor (\lambda >0\land -\frac{\lambda }{4}<Q<0$ & \\
 & & $\land \beta _{1,c}\leq \beta <\beta _{2,c}\land \sigma >0)$ & \\[15pt]
$\ \ \ \ \ \ \ \ d_M \ \ \ \ \ \ \ \ $ & $\left(Q<0\land \beta >\frac{1}{6} \left(2 Q^2-3\right)\right)$ & $(Q<-\frac{1}{\sqrt{2}}\land ((\beta <-\frac{1}{2}\land \lambda <\lambda _{d,2}$ & $\left(Q<0\land \beta_{d,2}<\beta <-\frac{1}{2}\right)$    \\
 & $\lor \left(Q>0\land \beta >\frac{1}{6} \left(2 Q^2-3\right)\right)$ & $\land \sigma <0)\lor (\beta >\beta _{d,1}\land \lambda >\lambda _{d,2}$ & $\lor \left(Q>0\land \beta_{d,2}<\beta <-\frac{1}{2}\right)$ \\
 & & $\land \sigma >0)))\lor (Q=-\frac{1}{\sqrt{2}}\land ((\beta <-\frac{1}{2}$ & & \\
 & & $\land \lambda <-\lambda _{d,1}\land \sigma <0)\lor (\beta >0\land \lambda >-\lambda _{d,1}$ & \\
 & & $\land \sigma >0)))\lor (-\frac{1}{\sqrt{2}}<Q<0\land ((\beta <-\frac{1}{2}$ & \\
 & & $\land \lambda <\lambda _{d,2}\land \sigma <0)\lor (\beta >\beta _{d,1}\land \lambda >\lambda _{d,2}$ & \\
 & & $\land \sigma >0)))\lor (0<Q<\frac{1}{\sqrt{2}}\land ((\beta <-\frac{1}{2}$ & \\
 & & $\land \lambda >\lambda _{d,2}\land \sigma >0)\lor (\beta >\beta _{d,1}\land \lambda <\lambda _{d,2}$ & \\
 & & $\land \sigma <0)))\lor (Q=\frac{1}{\sqrt{2}}\land ((\beta <-\frac{1}{2}\land \lambda >\lambda _{d,1}$ & \\
 & & $\land \sigma >0)\lor (\beta >0\land \lambda <\lambda _{d,1}\land \sigma <0)))$ & \\
 & & $\lor (Q>\frac{1}{\sqrt{2}}\land ((\beta <-\frac{1}{2}\land \lambda >\lambda _{d,2}\land \sigma >0)$ & \\
 & & $\lor (\beta >\beta _{d,1}\land \lambda <\lambda _{d,2}\land \sigma <0)))$ & \\[15pt]
$\ \ \ \ \ \ \ \ e^{\pm} \ \ \ \ \ \ \ \ $ & $\beta >-\frac{1}{2}$ & unstable for & never     \\
 & & $(Q>0\land ((\beta >-\frac{1}{2}\land \lambda <0\land \sigma <0)$ & \\
 & & $\lor (\lambda >0\land \beta >\beta _{1,e}\land \sigma <0)))\lor (Q<0$ & \\
 & & $\land \beta >\beta _{2,e}\land \lambda <0\land \sigma <0)\lor (\lambda >0$ & \\
 & & $\land ((Q\leq -\frac{\lambda }{2}\land \beta >\beta _{2,e})\lor (-\frac{\lambda }{2}<Q<0$ & \\
 & & $\land \beta >\beta _{1,e}))\land \sigma <0)$ & \\[15pt]
$\ \ \ \ \ \ \ \ f_{R} \ \ \ \ \ \ \ \ $ & $\left(\lambda <0\land -\frac{1}{2}<\beta <\frac{1}{8} \left(\lambda ^2-4\right)\right)$ & $(\lambda <0\land Q>-\frac{\lambda }{4}\land \beta _{f,1}\leq \beta <\beta _{f,2}\land \sigma <0)$ & never  \\
 & $\lor \left(\lambda >0\land -\frac{1}{2}<\beta <\frac{1}{8} \left(\lambda ^2-4\right)\right)$ & $\lor \left(\lambda >0\land Q<-\frac{\lambda }{4}\land \beta _{f,1}\leq \beta <\beta _{f,2}\land \sigma >0\right)$ & \\[15pt]
 $\ \ \ \ \ \ \ \ g \ \ \ \ \ \ \ \ $ & $\left(\lambda <0\land \left(\beta <-\frac{1}{2}\lor \beta \geq \beta _{1,g}\right)\right)$ & $(\lambda <0\land ((\beta <-\frac{1}{2}\land Q<Q_{1,g}\land \sigma >0)$ & $\left(\lambda <0\land \left(\beta <-\frac{1}{2}\lor \beta >\beta_{2,g}\right)\right)$  \\
 & $\lor \left(\lambda >0\land \left(\beta <-\frac{1}{2}\lor \beta \geq \beta _{1,g}\right)\right)$ & $\lor (\beta >\frac{1}{8} \left(\lambda ^2-4\right)\land Q>Q_{1,g}\land \sigma <0)))$ & $\lor \left(\lambda >0\land \left(\beta <-\frac{1}{2}\lor \beta >\beta_{2,g}\right)\right)$ \\
 & & $\lor (\lambda >0\land ((\beta <-\frac{1}{2}\land Q>Q_{1,g}\land \sigma <0)$ &  \\
 & & $\lor (\beta >\frac{1}{8} (\lambda ^2-4)\land Q<Q_{1,g}\land \sigma >0)))$ & \\[15pt]
$\ \ \ \ \ \ \ \ h \ \ \ \ \ \ \ \ $ & $(\lambda <0\land ((Q<0\land \beta \geq \beta _{h,0})$ & $(\lambda <0\land ((Q<0\land ((\beta _{h,1}\leq \beta <\beta _{h,2}\land \sigma <0)$ & $\left(\lambda <0\land \left(Q<\frac{\lambda }{2}\lor Q>-\lambda \right)\right)$  \\
 & $\lor (0<Q<-\lambda \land \beta \geq \beta _{h,0}) $ & $\lor (\beta _{h,3}\leq \beta <\beta _{h,4}\land \sigma <0)))\lor (0<Q<-\frac{\lambda }{4}$ & $\lor \left(\lambda >0\land \left(Q<-\lambda \lor Q>\frac{\lambda }{2}\right)\right)$\\
 & $\lor (Q>-\lambda\land \beta \geq \beta _{h,0})))\lor (\lambda >0$ & $\land ((\beta _{h,2}<\beta \leq \beta _{h,1}\land \sigma <0)$ & \\
 & $\land ((Q<-\lambda \land \beta \geq \beta _{h,0})\lor (-\lambda <Q<0$ & $\lor (\beta _{h,3}\leq \beta <\beta _{h,4}\land \sigma <0)))$ & \\
 & $\land \beta \geq \beta _{h,0})\lor (Q>0\land \beta \geq \beta _{h,0})))$ & $\lor (Q>-\lambda \land ((\beta _{h,1}\leq \beta <\beta _{h,2}\land \sigma >0)$ & \\
 & & $\lor (\beta _{h,4}<\beta \leq \beta _{h,3}\land \sigma >0)))))\lor (\lambda >0$ & \\
 & & $\land ((Q<-\lambda \land ((\beta _{h,1}\leq \beta <\beta _{h,2}\land \sigma <0)$ & \\
 & & $\lor (\beta _{h,4}<\beta \leq \beta _{h,3}\land \sigma <0)))\lor (-\frac{\lambda }{4}<Q<0$ & \\
 & & $\land ((\beta _{h,2}<\beta \leq \beta _{h,1}\land \sigma >0)\lor (\beta _{h,3}\leq \beta <\beta _{h,4}$ & \\
 & & $\land \sigma >0)))\lor (Q>0\land ((\beta _{h,1}\leq \beta <\beta _{h,2}$ & \\
 & & $\land \sigma >0)\lor (\beta _{h,3}\leq \beta <\beta _{h,4}\land \sigma >0)))))$ & \\[15pt]
$\ \ \ \ \ \ \ \ i_M \ \ \ \ \ \ \ \ $ & $(\sigma <0\land Q>0)\lor (\sigma >0\land Q<0)$ & $(\sigma <0\land ((\beta <-\frac{1}{2}\land 0<Q\leq \frac{6 \beta +3}{16 \sigma })$ & never   \\
& & $\lor (\beta >-\frac{1}{2}\land \frac{6 \beta +3}{16 \sigma }\leq Q<0)))\lor (\sigma >0$ & \\
& & $\land ((\beta <-\frac{1}{2}\land \frac{6 \beta +3}{16 \sigma }\leq Q<0)\lor (\beta >-\frac{1}{2}$ & \\
& & $\land 0<Q\leq \frac{6 \beta +3}{16 \sigma })))$ & \\[15pt]

\hline\hline
\end{tabular}
\end{center}
\label{tab:A1}
\end{table*}

\hfill \break
\hfill \break
\hfill \break
\hfill \break
\hfill \break
\hfill \break
\hfill \break

\section{Hubble's rate analysis}\label{appen_H}

For this analysis, we utilize a dataset consisting of 39 data points for $0.01 < z < 2.360$, as outlined in Table \ref{table:H(z)data}. 


\begin{table}[!b]
\caption{Hubble's parameter vs. redshift \& scale factor.}
\label{table:H(z)data}
\renewcommand{\tabcolsep}{0.7pc} 
\renewcommand{\arraystretch}{0.7} 
\begin{tabular}{@{}lllll}
\hline \hline
  $\;\; z$    &  $ H(z) \;$ ($\frac{km/s}{\text{Mpc}}$ ) &  Ref. \\
\hline
$0.07$      & $ \; \qquad 69     \pm 19.6 $      &   \cite{zhang2014} \\
$0.09$      & $ \; \qquad 69     \pm 12 $      & \cite{simon2005} \\
$0.100$     & $ \; \qquad 69     \pm 12 $      & \cite{simon2005} \\
$0.120$     & $ \; \qquad 68.6     \pm 26.2$       & \cite{zhang2014} \\
$0.170$     & $ \; \qquad 83     \pm 8$       & \cite{simon2005} \\
$0.179$     & $ \; \qquad 75     \pm 4$       & \cite{moresco2012} \\
$0.199$     & $ \; \qquad 75     \pm 5$        & \cite{moresco2012} \\
$0.200$     & $ \; \qquad 72.9     \pm 29.6$        &  \cite{zhang2014} \\
$0.270$     & $ \; \qquad 77     \pm 14$      & \cite{simon2005} \\
$0.280$     & $ \; \qquad 88.8     \pm 36.6$      & \cite{zhang2014} \\
$0.320$     & $ \; \qquad 79.2   \pm 5.6$     & \cite{cuesta2016}\\
$0.352$     & $ \; \qquad 83     \pm 14$      & \cite{moresco2012} \\
$0.3802$    & $ \; \qquad 83     \pm 13.5$      & \cite{moresco2012} \\
$0.400$     & $ \; \qquad 95     \pm 17$      & \cite{simon2005} \\
$0.4004$    & $ \; \qquad 77     \pm 10.2$      & \cite{moresco2012} \\
$0.4247$    & $ \; \qquad 87.1     \pm 11.2$      & \cite{moresco2012} \\
$0.440$     & $ \; \qquad 82.6   \pm 7.8$     & \cite{blake2012} \\
$0.4497$    & $ \; \qquad 92.8   \pm 12.9$     & \cite{moresco2012} \\
$0.470$     & $ \; \qquad 89   \pm 50$     & \cite{ratsim} \\
$0.4783$    & $ \; \qquad 80.9   \pm 9$     & \cite{moresco2012} \\
$0.480$     & $ \; \qquad 97     \pm 62$      & \cite{stern2010} \\
$0.570$     & $ \; \qquad 100.3  \pm 3.7$     & \cite{cuesta2016} \\
$0.593$     & $ \; \qquad  104   \pm 13$      & \cite{moresco2012} \\
$0.600$     & $ \; \qquad 87.9   \pm 6.1$     & \cite{blake2012} \\
$0.680$     & $ \; \qquad 92     \pm 8$       & \cite{moresco2012} \\
$0.730$     & $ \; \qquad 97.3   \pm 7 $      & \cite{blake2012} \\
$0.781$     & $ \; \qquad 105    \pm 12$      & \cite{moresco2012} \\
$0.875$     & $ \; \qquad 125    \pm 17$      & \cite{moresco2012} \\
$0.880$     & $ \; \qquad 90     \pm 40$      & \cite{stern2010} \\
$0.900$     & $ \; \qquad 117    \pm 23$      & \cite{simon2005} \\
$1.037$     & $ \; \qquad 154    \pm 20 $     & \cite{moresco2012} \\
$1.300$     & $ \; \qquad 168    \pm 17 $     & \cite{simon2005} \\
$1.363$     & $ \; \qquad 160    \pm 33.6$    & \cite{moresco2015}\\
$1.430$     & $ \; \qquad 177    \pm 18$      & \cite{simon2005}\\
$1.530$     & $ \; \qquad 140    \pm 14$      & \cite{simon2005}\\
$1.750$     & $ \; \qquad 202    \pm 40$      & \cite{simon2005}\\
$1.965$     & $ \; \qquad 186.5  \pm 50.4$    & \cite{moresco2015}\\
$2.340$     & $ \; \qquad 222    \pm 7 $      & \cite{delubac2014}\\
$2.360$     & $ \; \qquad 226    \pm  8$      & \cite{font-ribera2014}\\
\hline
\end{tabular}\\
 \end{table}
 
\end{appendix}

\end{document}